\documentclass[11pt,a4paper]{article}
\usepackage{jcappub}

\usepackage{bm,Macro,tabularx}

\def\uC{C}

\def\uTT{\uC^{\Theta\Theta}}
\def\uTE{\uC^{\Theta E}}
\def\uEE{\uC^{EE}}
\def\uBB{\uC^{BB}}

\def\grad{\phi}
\def\curl{\varpi}

\def\wig{Wigner-3{\it j} symbols }
\def\PL{PLANCK }

\def\AP{ACTPol }

\def\ol{\overline}

\def\bl{\bm{\ell}}
\def\bL{\bm{L}}
\newcommand\Int[1]{\int\frac{d^2 #1}{(2\pi)^2}}
\def\btbc{\bc \begin{table}[t]}
\def\etec{\end{table}\ec}
\def\bfbc{\begin{figure}[t] \bc}
\def\efec{\ec \end{figure}}

\title{Full-sky lensing reconstruction of gradient and curl modes from CMB maps}

\author[a]{Toshiya Namikawa,}
\author[b]{Daisuke Yamauchi}
\author[c,d]{and Atsushi Taruya}

\affiliation[a]{Department of Physics, Graduate School of Science, The University of Tokyo \\ 
Tokyo 113-0033, Japan}
\affiliation[b]{Institute for Cosmic Ray Research, The University of Tokyo \\
5-1-5 Kashiwa-no-ha, Kashiwa City, Chiba 277-8582, Japan}
\affiliation[c]{Research Center for the Early Universe, School of Science, The University of Tokyo \\ 
Bunkyo-ku, Tokyo 113-0033, Japan}
\affiliation[d]{Institute for the Physics and Mathematics of the Universe, The University of Tokyo \\ 
Kashiwa, Chiba 277-8568, Japan}

\emailAdd{namikawa@utap.phys.s.u-tokyo.ac.jp}
\emailAdd{yamauchi@icrr.u-tokyo.ac.jp}
\emailAdd{ataruya@utap.phys.s.u-tokyo.ac.jp}

\abstract{
We present a method of lensing reconstruction on the full sky, by extending the optimal quadratic 
estimator proposed by Okamoto \& Hu (2003) to the case including the curl mode of deflection 
angle. The curl mode is induced by the vector and tensor metric perturbations, and the reconstruction 
of the curl mode would be a powerful tool to not only check systematics in the estimated gradient 
mode but also probe any vector and tensor sources. We find that the gradient and curl modes can 
be reconstructed separately, thanks to the distinctive feature in the parity symmetry between the 
gradient and curl modes. We compare our estimator with the flat-sky estimator proposed by Cooray 
{\it et al} (2005). Based on the new formalism, the expected signal-to-noise ratio of the curl mode 
produced by the primordial gravitational-waves and a specific model of cosmic strings are estimated, 
and prospects for future observations are discussed. 
}

\begin{document}
\maketitle 

\section{
Introduction
} 
\label{sec.1}

Ongoing, upcoming and next-generation experiments of CMB, such as 
PLANCK \cite{PLANCK}, POLARBEAR \cite{POLARBEAR}, ACTPol \cite{Niemack:2010wz}, 
SPTPol \cite{SPTPol}, CMBPol \cite{CMBPol}, and COrE \cite{COrE} would 
resolve not only the fine structure of 
the primary CMB anisotropies but also tiny effects that are important at 
small scales. One of the most important signals in those effects is the 
weak lensing: the deflection of CMB photons coming from the last-scattering 
surface by metric perturbations along our line-of-sight. Recent studies 
show that the lensing fields involved in the CMB anisotropies are 
reconstructed from the cross-correlations between CMB and large-scale 
structure \cite{Smith07,Hirata:2008cb}, and CMB maps alone 
\cite{Smidt:2010by,Das:2011ak,Sherwin:2011gv}, and the lensing 
information from upcoming experiments would provide us an opportunity to 
probe the late-time evolution of the structure in the Universe 
(e.g., \cite{Hu:2001fb,Kaplinghat:2003bh,Lesgourgues:2006nd,
Das:2008am,Calabrese:2008rt,dP09,HSCS09,Namikawa:2010re}). 

Several studies have investigated a method to reconstruct the deflection 
angle, which characterizes the effect of weak lensing on CMB maps 
(e.g., \cite{Bernardeau:1998mw,Zaldarriaga:1998te,Seljak:1998aq,Hu:2001tn,Hu:2001kj,
Okamoto:2003zw,Hirata:2003ka,Carvalho:2010rz,Anderes:2010fq}). In 
general, the deflection angle in a direction $\hatn$, $\bm{d}(\hatn)$, 
where $\hatn$ is the unit vector defined on the unit sphere, is 
decomposed into gradient and curl part as (e.g., \cite{Hirata:2003ka}) 
\beq 
  \bm{d}(\hatn) 
    = \bm{\nabla}\grad(\hatn) + (\star \bm{\nabla})\curl(\hatn) 
  \,,
  \label{deflection}
\eeq 
where the first term, $\bm{\nabla}\grad(\hatn)$, and second term, 
$(\star\bm{\nabla})\curl(\hatn)$, represent gradient and curl mode of 
deflection angle, respectively, and, in the polar coordinate, the 
covariant derivative on the unit sphere is given by 
$\bm{\nabla}=\bm{e}_{\theta}(\pd/\pd\theta)+(\bm{e}_{\varphi}/\sin\theta)(\pd/\pd\varphi)$ 
with $\bm{e}_{\theta}$ and $\bm{e}_{\varphi}$ describing the basis 
vectors in the polar coordinate. The symbol, $\star$, denotes a operation 
which rotates the angle of two-dimensional vector counterclockwise by 
90-degree; for a vector on the unit sphere expressed in terms of the 
basis vectors, 
$\bm{a}=a_{\theta}\bm{e}_{\theta}+a_{\varphi}\bm{e}_{\varphi}$, the 
operator, $\star$, act on $\bm{a}$ as 
$(\star\bm{a})=a_{\theta}\bm{e}_{\varphi}-a_{\varphi}\bm{e}_{\theta}$ 
\cite{Hirata:2003ka}. Hereafter, we call the potentials, $\grad$ and 
$\curl$, ``scalar lensing potential'' and ``pseudo-scalar lensing potential'', 
respectively. 

The scalar metric perturbations such as the matter density fluctuations at 
linear order produce only the gradient mode, and the curl mode is usually 
neglected in the algorithm of lensing reconstruction. However, the curl 
mode can be induced by vector and/or tensor metric perturbations. In the 
conformal Newton gauge, the line element in the polar coordinate system 
is given as 
\beq 
  ds^2 = a^2(\eta)\{-(1+2A)d\eta^2 - 2B_id\eta dx^i 
    + [(1+2C)\gamma_{ij}+2D_{ij}]dx^i dx^j\} 
  \,, 
  \label{ds-P}
\eeq 
where $a$ is the scale factor, $\eta$ is the conformal time, $A$ and $C$ 
are the scalar components, $B_i$ is the vector component ($B_{i|i}=0$), 
and $D_{ij}$ is the tensor component ($D_{ij|i}=0$, and $D_{ii}=0$). 
The unperturbed spatial metric, $\gamma_{ij}$, is given by 
\beq 
  \gamma_{ij}dx^idx^j = d\chi^2 + \chi^2(d\theta^2+\sin^2\theta d\varphi^2)
  \,. 
\eeq 
Then, the pseudo-scalar lensing potential is described by \cite{Yamauchi:2011}
\al{ 
  \curl(\hatn) 
    &= (\star\bm{\nabla})^{-2}\int_0^{\chi_s}\frac{d\chi}{\chi^2\sin\theta}
      \left[\PD{\Omega_{\theta}(\eta_0-\chi,\chi\hatn)}{\varphi}
      - \PD{\Omega_{\varphi}(\eta_0-\chi,\chi\hatn)}{\theta}\right]
  \,,
} 
where the quantities, $\chi$ and $\chi_s$, are the comoving distance and 
comoving distance at the last scattering surface, respectively, $\eta_0$ 
is the conformal time today, and the quantities, $\Omega_{\theta}$ and 
$\Omega_{\varphi}$, are defined as 
\beq 
  \Omega_a(\eta,\chi\hatn) = B_a(\eta,\chi\hatn) + 2D_{\chi a}(\eta,\chi\hatn) 
  \qquad 
  (a=\theta,\varphi)
  \,. 
\eeq 
According to the above equations, the primordial gravitational-waves 
produce the curl mode (e.g.,\cite{Li:2006si}). The cosmic strings can 
also produce the curl mode through the vector and tensor perturbations 
(e.g.,\cite{Benabed:1999wn,Yamauchi:2011}). This implies that the curl 
mode is a smoking gun of cosmic strings and other vector or tensor 
sources. Even if these sources are absent, the higher-order density 
perturbations and foreground contaminations generate not only the 
gradient mode but also the curl mode. These sources would cause 
systematics in the estimation of the gradient mode. Thus, the evaluation 
of the contaminations of these sources in the curl mode would be helpful 
to estimate the contributions of these sources in the gradient mode. 

In this paper, we present a method to reconstruct both the gradient and 
curl modes of deflection angle. The reconstruction of the curl mode has 
been previously discussed in Refs.\cite{Hirata:2003ka,Cooray:2005hm}. 
Ref.\cite{Hirata:2003ka} proposed an algorithm based on the likelihood 
analysis. Including polarizations, this estimator, in principle, 
suppresses the noise contribution for the reconstructed potentials, but 
numerically cost, compared to the estimator in Ref.\cite{Hu:2001kj}. 
On the other hand, in Ref.\cite{Cooray:2005hm}, they consider the 
flat-sky limit and empirically define a quadratic estimator of the curl 
mode. Since current and future CMB missions will cover nearly entire sky, 
a full-sky algorithm including the curl mode are highly desirable. 
In this paper, we derive a full-sky estimator of the gradient and curl 
modes on the full sky, extending the full-sky formalism for the gradient 
mode in Ref.\cite{Okamoto:2003zw}. Then, we compare the flat-sky 
estimator with the one on the full sky, and show that the empirically 
defined estimator in Ref.\cite{Cooray:2005hm} can be derived from the 
full-sky estimator. In addition, based on our full-sky estimator, 
possible implications to detection of curl mode from primordial 
gravitational-waves and cosmic strings are discussed. 

This paper is organized as follows. In section \ref{sec.2}, we briefly 
summarize the lensing effect on the CMB anisotropies. In section 
\ref{sec.3}, we extend the quadratic estimator of 
Ref.\cite{Okamoto:2003zw} to the case including both the gradient and 
curl modes of deflection angle. In section \ref{sec.4}, we compute the 
noise spectrum of the estimator in both the full- and flat-sky cases. 
In section \ref{sec.5}, we discuss implications for primordial 
gravitational-waves and cosmic strings by reconstructing the curl mode. 
Section \ref{sec.6} is devoted to summary and conclusion. 

In this paper we adopt the cosmological parameters assuming a flat 
Lambda-CDM model consistent with the results obtained from 
Ref.\cite{Dunkley:2010ge}; the density parameter of baryon 
$\Omega\rom{b}h^2=0.022$, of matter $\Omega\rom{m}h^2=0.13$, dark energy 
density $\Omega_{\Lambda}=0.72$, scalar spectral index $n\rom{s}=0.96$, 
scalar amplitude $A\rom{s}=2.4\times10^{-9}$ and the optical depth, 
$\tau=0.086$. In Table.\ref{notation}, we summarize the meaning and 
definition of the quantities used to reconstruct the lensing potentials 
from observed CMB maps. 

\btbc
\caption{Notations for quantities used to reconstruct the lensing 
potentials from observed CMB maps. The symbols are divided into three 
categories. The quantities in the middle eight rows are needed to compute 
the optimal quadratic estimator. In the bottom three rows, we describe 
the quantities needed to compute the optimal combination.} 
\vs{0.5}
\begin{tabularx}{14.5cm}{clX} \hline \hline
Symbol & \multicolumn{1}{c}{Definition} & Meaning \\ \hline 
Full sky / Flat sky & Full sky / Flat sky & \\ \hline \hline
$x$ ($=\grad$, $\curl$) & Eq.(\ref{deflection}) & Scalar or pseudo-scalar lensing potential \\ 
$\alpha$ (or $\beta$) & - & A pair of two CMB maps, $X$ and $Y$ \\ \hline 
$\hat{x}_{\ell,m}^{(\alpha)}$ / $\hat{x}_{\bl}^{(\alpha)}$ & 
Eq.(\ref{3:est}) / (\ref{flat:est}) & Estimator \\ 
$F^{x,(\alpha)}_{\ell,L,L\p}$ / $\ol{F}^{x,(\alpha)}_{\bl,\bL,\bL\p}$ & 
Eq.(\ref{3:F}) / (\ref{flat:F}) & Weight function \\ 
$N^{x,(\alpha)}_{\ell}$ / $\ol{N}^{x,(\alpha)}_{\ell}$ & 
Eq.(\ref{3:N}) / (\ref{flat:N}) & Noise spectrum \\  
$g^{x,(\alpha)}_{\ell,L,L\p}$ / $\ol{g}^{x,(\alpha)}_{\bl,\bL,\bL\p}$ & 
Eq.(\ref{full:g-func}) / (\ref{flat:g-func}) & - \\ 
$f^{x,(\alpha)}_{\ell,L,L\p}$ / $\ol{f}^{x,(\alpha)}_{\bl,\bL,\bL\p}$ & 
Table.\ref{table:f} / \ref{table:olf} & - \\ 
$\,_0\mcS^{x}$ & Eq.(\ref{0mcS}) & - \\ 
$\,_{\pm2}\mcS^{x}$ & Eq.(\ref{2mcS}) & - \\ 
$\,_{\oplus}\mcS^{x}$ and $\,_{\ominus}\mcS^{x}$ & Eq.(\ref{pmmcS}) & - \\ \hline 
$\hat{x}_{\ell,m}^{({\rm c})}$ / $\hat{x}_{\ell,m}^{({\rm c})}$ & 
Eq.(\ref{3:c}) / (\ref{flat:c}) & Optimal combination \\ 
$N^{x,({\rm c})}_{\ell}$ / $\ol{N}^{x,({\rm c})}_{\ell}$ & 
Eq.(\ref{3:N-c}) / (\ref{flat:N-c}) & Noise spectrum for optimal combination \\ 
$N^{x,(\alpha,\beta)}_{\ell}$ / $\ol{N}^{x,(\alpha,\beta)}_{\ell}$ & 
Eq.(\ref{3:covar}) / (\ref{flat:covar}) & Noise cross-spectrum \\ \hline 
\end{tabularx}
\label{notation}
\etec

\section{
Weak lensing of the CMB
} 
\label{sec.2} 

Here we briefly review the lensing effect on the CMB anisotropies in the 
full-sky case, including both the gradient and curl modes of deflection 
angle. We first discuss the lensing effect on CMB temperature 
in section \ref{sec.2.1}, and the similar discussion for polarizations 
is given in section \ref{sec.2.2}. The detailed calculation of lensing 
effect is presented in, e.g., Ref.\cite{Hu:2000ee} in the absence of 
the curl mode, and in Ref.\cite{Li:2006si} including the curl mode. 

\subsection{
Lensing effect on CMB anisotropies: temperature
} 
\label{sec.2.1}

Let us first discuss the lensing effect on the temperature anisotropies. 
The lensed temperature fluctuations in a direction $\hatn$, 
$\tilde{\Theta}(\hatn)$, are transformed into the harmonic space according to 
\beq 
  \tilde{\Theta}_{\ell,m} 
    = \int d\hatn \,_0\mcY_{\ell,m}^*(\hatn)\tilde{\Theta}(\hatn)
  \,, 
  \label{harmonic}
\eeq 
with the quantities, $\tilde{\Theta}_{\ell,m}$ and 
$\,_0\mcY_{\ell,m}(\hatn)$, describing the harmonic coefficients, and the 
spin-$0$ spherical harmonics, respectively. 
The lensed temperature fluctuations are related to the unlensed 
temperature fluctuations, $\Theta(\hatn)$, through 
$\tilde{\Theta}(\hatn) = \Theta(\hatn+\bm{d})$, where $\bm{d}$ is the 
deflection angle. Usually, the deflection angle is a small perturbed 
quantity, $|\bm{d}|\ll 1$, and the lensed 
temperature fluctuations may be expressed as  
\beq 
  \tilde{\Theta}(\hatn) 
    = \Theta(\hatn) + \bm{d}\cdot\bm{\nabla}\Theta(\hatn) 
    + \mcO(|\bm{d}|^2)
  \,, 
  \label{remap} 
\eeq 
where we expand $\Theta(\hatn+\bm{d})$ in terms of the deflection angle 
$\bm{d}$. Hereafter we neglect the contributions 
of $\mcO(|\bm{d}|^2)$ in Eq.(\ref{remap}). The harmonic coefficients 
of the lensed quantities are obtained by transforming Eq.(\ref{remap}) 
into the harmonic space, according to Eq.(\ref{harmonic}). Using the 
expression of deflection angle (\ref{deflection}), the lensed 
temperature anisotropies in the harmonic space are given by 
\cite{Li:2006si} 
\al{ 
  \tilde{\Theta}_{L,M} 
  &= \Theta_{L,M} 
    + \int d\hatn \,_0 \mcY_{L,M}^*(\hatn)
    \big[\bm{\nabla}\grad(\hatn) + (\star\bm{\nabla})\curl(\hatn)\big]\cdot
    \bm{\nabla}\Theta(\hatn) 
  \notag 
  \\ 
  &= \Theta_{L,M} + \sum_{\ell,m}\sum_{\ell\p,m\p}
    \Theta_{\ell\p,m\p}(-1)^M\Wjm{L}{\ell}{\ell\p}{-M}{m}{m\p}
    \sum_{x=\grad,\curl}\,_0\mcS^x_{L,\ell,\ell\p}x_{\ell,m} 
  \,,
  \label{lens:T}
} 
where the quantities, $\,_0\mcS^{\grad}_{L,\ell,\ell\p}$ and 
$\,_0\mcS^{\curl}_{L,\ell,\ell\p}$, are defined as 
\al{
  (-1)^M\Wjm{L}{\ell}{\ell\p}{-M}{m}{m\p}\,_0 \mcS^{\grad}_{L,\ell,\ell\p}
    &= \int d\hatn \,_0\mcY_{L,M}^*(\hatn)
    [\bm{\nabla}\,_0\mcY_{\ell,m}(\hatn)]\cdot
    [\bm{\nabla}\,_0\mcY_{\ell\p,m\p}(\hatn)] 
  \,, 
  \label{mcSg} 
  \\ 
  (-1)^M\Wjm{L}{\ell}{\ell\p}{-M}{m}{m\p}\,_0 \mcS^{\curl}_{L,\ell,\ell\p} 
    &= \int d\hatn \,_0\mcY_{L,M}^*(\hatn)
    [(\star\bm{\nabla})\,_0\mcY_{\ell,m}(\hatn)]\cdot
    [\bm{\nabla}\,_0\mcY_{\ell\p,m\p}(\hatn)] 
  \,.
  \label{mcSc}
}
As shown in appendix \ref{appA} (Eqs.(\ref{Wigner:harmonics1}) and 
(\ref{Wigner:harmonics2})), the quantities, $\,_0\mcS^{\grad}_{L,\ell,\ell\p}$ and 
$\,_0\mcS^{\curl}_{L,\ell,\ell\p}$, are expressed in terms of the Wigner-3$j$ symbols, and 
the results are 
\al{ 
  \,_0 \mcS^{\grad}_{L,\ell,\ell\p} 
    &= \sqrt{\frac{(2\ell+1)(2\ell\p+1)(2L+1)}{16\pi}}
    [-L(L+1)+\ell(\ell+1)+\ell\p(\ell\p+1)]
    \Wjm{L}{\ell}{\ell\p}{0}{0}{0} 
  \,,
  \\ 
  \,_0 \mcS^{\curl}_{L,\ell,\ell\p}
    &= -i\sqrt{\frac{(2\ell+1)(2\ell\p+1)(2L+1)}{16\pi}}
    \sqrt{\ell(\ell+1)}\sqrt{\ell\p(\ell\p+1)} \notag \\ 
    &\qquad\qquad \times 
    \left[\Wjm{L}{\ell}{\ell\p}{0}{-1}{1}-\Wjm{L}{\ell}{\ell\p}{0}{1}{-1}\right]
  \,.
  \label{0mcS}
} 
We note that the quantities $\,_0 \mcS^{\grad}_{L,\ell,\ell\p}$ and 
$\,_0 \mcS^{\curl}_{L,\ell,\ell\p}$ satisfy 
\al{ 
  \,_0 \mcS^{\grad}_{L,\ell,\ell\p} 
    &= (-1)^{L+\ell+\ell\p}\,_0 \mcS^{\grad}_{L,\ell,\ell\p}
  \,, 
  &
  \,_0 \mcS^{\curl}_{L,\ell,\ell\p} 
    &= -(-1)^{L+\ell+\ell\p}\,_0 \mcS^{\curl}_{L,\ell,\ell\p}
  \,. 
  \label{F:parity}
}
The above equations come from the parity symmetry of $\Theta$, $\grad$ 
and $\curl$; the temperature anisotropies and the scalar lensing 
potential are even parity, while the pseudo-scalar lensing potential 
is odd parity. In fact, Eq.(\ref{F:parity}) is checked by changing the 
variable, $\hatn\to-\hatn$ in the right-hand side of Eqs.(\ref{mcSg}) 
and (\ref{mcSc}). Under this transformation, the spin-$0$ spherical 
harmonics are multiplied by a factor $(-1)^{\ell}$, and the derivatives 
become $\bm{\nabla}\to\bm{\nabla}$ and 
$(\star\bm{\nabla})\to -(\star\bm{\nabla})$, respectively. As a result, 
the right-hand sides of Eq.(\ref{mcSg}) and Eq.(\ref{mcSc}) are 
multiplied by a factor of $(-1)^{L+\ell+\ell\p}$ and 
$-(-1)^{L+\ell+\ell\p}$, respectively. Eq.(\ref{F:parity}) is also 
checked with the formulas of the \wig (see Eq.(\ref{sym}) in appendix \ref{appA}). 

From Eq.(\ref{F:parity}), $\,_0 \mcS^{\grad}_{L,\ell,\ell\p}$ becomes 
zero if $L+\ell+\ell\p$ is an odd integer, and the coefficient 
$\,_0 \mcS^{\curl}_{L,\ell,\ell\p}$ vanishes when $L+\ell+\ell\p$ is an 
even integer. These properties are essential for a separate reconstruction 
of gradient and curl modes in subsequent analysis.

\subsection{
Lensing effect on CMB anisotropies: polarizations
} 
\label{sec.2.2}

Next we consider the lensing effect on the CMB polarizations, 
$Q(\hatn)\pm iU(\hatn)$. We are especially concerned with the 
rotationally invariant combinations, i.e., $E$- and $B$-mode 
polarizations \cite{Li:2006si}; 
\al{ 
  [\tilde{E}\pm i\tilde{B}]_{L,M} 
    &= [E\pm iB]_{L,M} 
      + \int d\hatn \,_{\pm 2} \mcY_{L,M}^*(\hatn)
      \big[\bm{\nabla}\grad(\hatn)+(\star\bm{\nabla})\curl(\hatn)\big]
      \bm{\nabla}(Q\pm iU)(\hatn) 
  \notag 
  \\ 
    &= [E\pm iB]_{L,M} 
      + \sum_{\ell,m}\sum_{\ell\p,m\p}[E\pm iB]_{\ell\p,m\p}
      (-1)^M\Wjm{L}{\ell}{\ell\p}{-M}{m}{m\p} \notag \\ 
      &\qquad\qquad\qquad\qquad\qquad\qquad\qquad\qquad \times 
      \sum_{x=\grad,\curl}\,_{\pm 2}\mcS^{x}_{L,\ell,\ell\p}x_{\ell,m}
  \,, 
  \label{lens:Poll}
} 
with the quantities, $\,_{\pm 2}\mcS^{\grad}_{L,\ell,\ell\p}$ and 
$\,_{\pm 2}\mcS^{\curl}_{L,\ell,\ell\p}$, defined by 
\al{
  (-1)^M\Wjm{L}{\ell}{\ell\p}{-M}{m}{m\p}\,_{\pm 2}\mcS^{\grad}_{L,\ell,\ell\p} 
    &= \int d\hatn \,_{\pm 2}\mcY_{L,M}^*(\hatn)
    [\bm{\nabla}\,_0\mcY_{\ell,m}(\hatn)]\cdot
    [\bm{\nabla}\,_{\pm 2}\mcY_{\ell\p,m\p}(\hatn)] 
  \,, 
  \label{2mcSg} 
  \\ 
  (-1)^M\Wjm{L}{\ell}{\ell\p}{-M}{m}{m\p}\,_{\pm 2}\mcS^{\curl}_{L,\ell,\ell\p} 
    &= \int d\hatn \,_{\pm 2}\mcY_{L,M}^*(\hatn)
    [(\star\bm{\nabla})\,_0\mcY_{\ell,m}(\hatn)]\cdot
    [\bm{\nabla}\,_{\pm 2}\mcY_{\ell\p,m\p}(\hatn)] 
  \,. 
  \label{2mcSc} 
}
The quantity $\,_{\pm 2}\mcY_{\ell\p,m\p}(\hatn)$ denotes the spin-$\pm 2$ 
spherical harmonics. Similar to the case of temperature anisotropies, 
the quantities, $\,_{\pm 2}\mcS^{\grad}_{L,\ell,\ell\p}$ and 
$\,_{\pm 2}\mcS^{\curl}_{L,\ell,\ell\p}$, are written as \cite{Li:2006si}
\al{ 
  \,_{\pm 2} \mcS^{\grad}_{L,\ell,\ell\p} 
    &= \sqrt{\frac{(2\ell+1)(2\ell\p+1)(2L+1)}{16\pi}}
    [\ell(\ell+1)+\ell\p(\ell\p+1)-L(L+1)]
    \Wjm{L}{\ell}{\ell\p}{\pm 2}{0}{\mp 2} 
  \,,
  \\ 
  \,_{\pm 2} \mcS^{\curl}_{L,\ell,\ell\p}
    &= -i\sqrt{\frac{(2\ell+1)(2\ell\p+1)(2L+1)}{16\pi}}
      \sqrt{\ell(\ell+1)}\sqrt{(\ell\p \pm 2)(\ell\p +1 \pm 2)} 
    \notag \\ 
    &\quad\times 
      \bigg[
        \sqrt{\frac{\ell\p +1 \mp 2}{\ell\p +1 \pm 2}}
        \Wjm{L}{\ell}{\ell\p}{\pm 2}{-1}{1 \mp 2} 
        - \sqrt{\frac{\ell\p \mp 2}{\ell\p \pm 2}}
        \Wjm{L}{\ell}{\ell\p}{\pm 2}{1}{-1 \mp 2}
      \bigg] 
  \,.
  \label{2mcS}
} 
Eq.(\ref{lens:Poll}) is rewritten in the separable form for 
$E$- and $B$-mode polarizations: 
\al{
  \tilde{E}_{L,M} &= E_{L,M} + \sum_{\ell,m}\sum_{\ell\p,m\p}(-1)^M
    \Wjm{L}{\ell}{\ell\p}{-M}{m}{m\p} \notag \\ 
    &\qquad\qquad\qquad \times
    \sum_{x=\grad,\curl}x_{\ell,m}
      \{\,_{\oplus }\mcS^{x}_{L,\ell,\ell\p} E_{\ell\p,m\p}
      - \,_{\ominus}\mcS^{x}_{L,\ell,\ell\p} B_{\ell\p,m\p}\} 
  \label{lens:E}
  \,,
  \\ 
  \tilde{B}_{L,M} &= B_{L,M} + \sum_{\ell,m}\sum_{\ell\p,m\p}(-1)^M
    \Wjm{L}{\ell}{\ell\p}{-M}{m}{m\p} \notag \\
    &\qquad\qquad\qquad \times
    \sum_{x=\grad,\curl}x_{\ell,m}
      \{\,_{\ominus}\mcS^{x}_{L,\ell,\ell\p} E_{\ell\p,m\p}
      + \,_{\oplus }\mcS^{x}_{L,\ell,\ell\p} B_{\ell\p,m\p}\} 
  \,, 
  \label{lens:B}
}
where we define 
\al{ 
  \,_{\oplus}\mcS^x_{L,\ell,\ell\p} 
    &= \frac{\,_{2}\mcS^x_{L,\ell,\ell\p}+\,_{-2}\mcS^x_{L,\ell,\ell\p}}{2} 
  \,, 
  & 
  \,_{\ominus}\mcS^x_{L,\ell,\ell\p} 
    &= \frac{\,_{2}\mcS^x_{L,\ell,\ell\p}-\,_{-2}\mcS^x_{L,\ell,\ell\p}}{2i} 
  \,. 
  \label{pmmcS}
} 
Note again that, for an even integer of $L+\ell+\ell\p$, the coefficients 
$\,{\oplus}\mcS^{\curl}$ and $\,_{\ominus}\mcS^{\grad}$ vanish. 
On the other hand, the 
quantities $\,_{\oplus}\mcS^{\grad}$ and $\,_{\ominus}\mcS^{\curl}$ vanish 
when $L+\ell+\ell\p$ is an odd integer. 
Similar to the case of temperature, these properties come 
from the fact that $E$-mode polarization and scalar lensing potential are 
even parity, while $B$-mode polarization and pseudo-scalar lensing potential 
are odd parity.

\section{
Reconstruction of deflection angle in the presence of curl mode
} 
\label{sec.3}

In this section, we present a reconstruction method for $\grad_{\ell,m}$ 
and $\curl_{\ell,m}$, based on the quadratic statistics (e.g., Refs.
\cite{Zaldarriaga:1998te,Hu:2001tn,Hu:2001kj,Okamoto:2003zw}). 
We frequently use the formulas for \wig summarized in appendix \ref{appA}. 
In what follows, the lensed temperature or polarizations, i.e., 
$\tilde{\Theta}$, $\tilde{E}$ or $\tilde{B}$, are symbolically denoted by 
$\tilde{X}$ (and $\tilde{Y}$).

\subsection{
Full-sky formalism
} 
\label{sec.3.1}

\subsubsection{
Lensing field as quadratic statistics
}

\btbc
\caption{The functional forms of 
$f^{\grad,(XY)}_{\ell,L,L^{\prime}} $ and 
$f^{\curl,(XY)}_{\ell,L,L^{\prime}} $ appeared in Eq.(\ref{3:XY}). 
The label ``even'' and ``odd'' indicate that the function are non-zero 
only when $\ell+L+L^{\prime}$ is even or odd, respectively.}
\vs{0.5}
\begin{tabular}{c|c|c} \hline 
$XY$ & $f^{\grad,(XY)}_{\ell,L,L\p}$ & $f^{\curl,(XY)}_{\ell,L,L\p}$ 
\\ \hline \\
$\Theta\Theta$ & 
$\,_0\mcS^{\grad}_{L,\ell,L\p}\uTT_{L\p}+\,_0\mcS^{\grad}_{L\p,\ell,L}\uTT_L$ (even) &
$\,_0\mcS^{\curl}_{L,\ell,L\p}\uTT_{L\p}-\,_0\mcS^{\curl}_{L\p,\ell,L}\uTT_L$ (odd), 
\\ \\ \hline \\ 
$\Theta E$ & 
$\,_0\mcS^{\grad}_{L,\ell,L\p}\uTE_{L\p}+\,_{\oplus}\mcS^{\grad}_{L\p,\ell,L}\uTE_L$ (even) &
$\,_0\mcS^{\curl}_{L,\ell,L\p}\uTE_{L\p}-\,_{\oplus}\mcS^{\curl}_{L\p,\ell,L}\uTE_L$ (odd) 
\\ \\ \hline \\ 
$\Theta B$ & 
$-\,_{\ominus}\mcS^{\grad}_{L\p,\ell,L}\uTE_L$ (odd) &
$\,_{\ominus}\mcS^{\curl}_{L\p,\ell,L}\uTE_L$ (even) 
\\ \\ \hline \\
$EE$ & 
$\,_{\oplus}\mcS^{\grad}_{L,\ell,L\p}\uEE_{L\p}+\,_{\oplus}\mcS^{\grad}_{L\p,\ell,L}\uEE_L$ (even) &
$\,_{\oplus}\mcS^{\curl}_{L,\ell,L\p}\uEE_{L\p}-\,_{\oplus}\mcS^{\curl}_{L\p,\ell,L}\uEE_L$ (odd) 
\\ \\ \hline \\ 
$EB$ & 
$-\,_{\ominus}\mcS^{\grad}_{L,\ell,L\p}\uBB_L-\,_{\ominus}\mcS^{\grad}_{L\p,\ell,L}\uEE_L$ (odd) & 
$-\,_{\ominus}\mcS^{\curl}_{L,\ell,L\p}\uBB_L+\,_{\ominus}\mcS^{\curl}_{L\p,\ell,L}\uEE_L$ (even) 
\\ \\ \hline \\
$BB$ & 
$\,_{\oplus}\mcS^{\grad}_{L,\ell,L\p}\uBB_{L\p}+\,_{\oplus}\mcS^{\grad}_{L\p,\ell,L}\uBB_L$ (even) &
$\,_{\oplus}\mcS^{\curl}_{L,\ell,L\p}\uBB_{L\p}-\,_{\oplus}\mcS^{\curl}_{L\p,\ell,L}\uBB_L$ (odd) 
\\ \\ \hline 
\end{tabular}
\label{table:f}
\etec

To get insight into the reconstruction of the lensing potentials, we 
consider an idealistic situation; we can take the ensemble average over 
primary CMB anisotropies alone, under a given realization of the lensing 
potentials. Hereafter, we denote this average by $\ave{\cdots}\rom{CMB}$, 
to distinguish it from the usual meaning of the ensemble average, 
$\ave{\cdots}$. Under the situation, in the correlation of lensed CMB 
anisotropies, 
$\langle \tilde{X}_{L,M}\tilde{Y}_{L\p,M\p}\rangle\rom{CMB}$, the lensing 
potentials are included in the non-zero off-diagonal terms 
($L\not=L\p$, $M\not=-M\p$). This is because the lensing effect causes a 
non-trivial mode-coupling between the primary CMB anisotropies and 
lensing potentials 
(see Eqs.(\ref{lens:T}), (\ref{lens:E}) and (\ref{lens:B})), 
and the lensed CMB anisotropies are not statistically isotropic for a 
given realization of the lensing potentials. Thus, it is possible to 
reconstruct the lensing potentials by extracting the off-diagonal terms 
of $\langle \tilde{X}_{L,M}\tilde{Y}_{L\p,M\p}\rangle\rom{CMB}$. 

To see this, let us write the correlation of the lensed CMB anisotropies 
in the harmonic space. 
With Eqs.(\ref{lens:T}), (\ref{lens:E}) and (\ref{lens:B}), we obtain 
\al{
  \ave{\tilde{X}_{L,M}\tilde{Y}_{L\p,M\p}}\rom{CMB} 
    &= C_L^{XY}\delta_{L,L\p}\delta_{M,-M\p}(-1)^M \notag \\
    &\qquad +\sum_{\ell,m}(-1)^m\Wjm{\ell}{L}{L\p}{-m}{M}{M\p}
      [f^{\grad,(XY)}_{\ell,L,L\p}\grad_{\ell,m} 
      + f^{\curl,(XY)}_{\ell,L,L\p}\curl_{\ell,m}] 
  \,, 
  \label{3:XY}
}
with the coefficients, $f^{\grad,(XY)}_{\ell,L,L\p}$ and 
$f^{\curl,(XY)}_{\ell,L,L\p}$, summarized in Table \ref{table:f}. 
The conditions, ``even'' and ``odd'', in Table \ref{table:f} 
come from the parity symmetry in the lensing potentials and primary CMB anisotropies. 
To extract the off-diagonal terms and find the solutions for 
$\grad_{\ell,m}$ and $\curl_{\ell,m}$, we multiply 
\al{
  (-1)^{m\p}\Wjm{\ell\p}{L}{L\p}{-m\p}{M}{M\p}f^{\grad,(XY)}_{\ell\p,L,L\p} 
  \,, 
  \label{3:multiply}
}
in both sides of Eq.(\ref{3:XY}). Note that the multipoles, $L$ and 
$L\p$, are chosen so that $f^{\grad,(XY)}_{\ell\p,L,L\p}\not=0$. 
Then, summing up the equation over $M$ and $M\p$, and using the formulas, 
Eqs.(\ref{Wigner:sum1}) and (\ref{Wigner:sum2}), we find
\footnote{
In deriving Eq.(\ref{3:grad-quad}), we have ignored the zero-mode 
($C_0^{XY}$), arising from the first term in Eq.(\ref{3:XY})
.} 
\al{
  \grad_{\ell,m} = 
    \frac{2\ell+1}{f^{\grad,(XY)}_{\ell,L,L\p}}
    \sum_{M,M\p}(-1)^m \Wjm{\ell}{L}{L\p}{-m}{M}{M\p}
    \ave{\tilde{X}_{L,M}\tilde{Y}_{L\p,M\p}}\rom{CMB} 
  \,. 
  \label{3:grad-quad}
}
Notice that the term involving $\curl$ in Eq.(\ref{3:XY}) vanishes. 
This is because, for all $\ell$, $L$ and $L\p$, the parity symmetry of 
$f^{\grad,(XY)}_{\ell,L,L\p}$ and $f^{\curl,(XY)}_{\ell,L,L\p}$ 
(Table \ref{table:f}) leads to 
\beq 
  f^{\grad,(XY)}_{\ell,L,L\p}f^{\curl,(XY)}_{\ell,L,L\p} = 0
  \,. 
  \label{3:sep}
\eeq 
Similarly, following the procedure described in Eq.(\ref{3:multiply}) 
below, but replacing $f^{\grad,(XY)}_{\ell,L,L\p}$ with $f^{\curl,(XY)}_{\ell,L,L\p}$, 
the solution for $\curl_{\ell,m}$ is obtained, and the result is 
\al{
  \curl_{\ell,m} = 
    \frac{2\ell+1}{f^{\curl,(XY)}_{\ell,L,L\p}}
    \sum_{M,M\p}(-1)^m \Wjm{\ell}{L}{L\p}{-m}{M}{M\p}
    \ave{\tilde{X}_{L,M}\tilde{Y}_{L\p,M\p}}\rom{CMB} 
  \,.
  \label{3:curl-quad}
}
The above equations, (\ref{3:grad-quad}) and (\ref{3:curl-quad}), 
can not be used for a definition of the estimator of lensing potentials, 
because these equations include the ensemble average over the primary 
CMB anisotropies alone, $\ave{\cdots}\rom{CMB}$. 
But, the above equations imply that, by summing the quadratic combination 
of lensed fields over multipoles appropriately, it is possible to 
separately construct the estimators for the scalar and pseudo-scalar 
lensing potentials, $\grad$ and $\curl$. 

\subsubsection{
Estimator
} 
\label{sec.3.1.2}

Based on Eqs.(\ref{3:grad-quad}) and (\ref{3:curl-quad}), we first 
naively define the estimator for the lensing potential $x$ 
($=\grad$ or $\curl$) as follows: 
\al{
  \frac{2\ell+1}{f^{x,(\alpha)}_{\ell,L,L\p}}
    \sum_{M,M\p}(-1)^m \Wjm{\ell}{L}{L\p}{-m}{M}{M\p}
    \tilde{X}_{L,M}\tilde{Y}_{L\p,M\p} 
  \,, 
  \label{3:est-naive}
}
where the subscript, $\alpha$, means a pair of two CMB maps, e.g., 
$\alpha=\Theta\Theta$ or $EB$. 
The multipoles, $L$ and $L\p$, are chosen so that 
$f^{x,(\alpha)}_{\ell,L,L\p}\not=0$. 
With Eqs.(\ref{3:grad-quad}) and (\ref{3:curl-quad}), the estimator is 
rewritten as 
\al{
  \text{Eq.(\ref{3:est-naive})} = x_{\ell,m} + n^{x,(\alpha)}_{\ell,m,L,L\p} 
  \,, 
  \label{3:expand}
}
where the quantity, $n^{x,(\alpha)}_{\ell,m,L,L\p}$, is given by 
\al{
  n^{x,(\alpha)}_{\ell,m,L,L\p} 
    &= (-1)^m \frac{2\ell+1}{f^{x,(\alpha)}_{\ell,L,L\p}}
      \sum_{M,M\p}\Wjm{\ell}{L}{L\p}{-m}{M}{M\p}
      \left(\tilde{X}_{L,M}\tilde{Y}_{L\p,M\p}
        -\ave{\tilde{X}_{L,M}\tilde{Y}_{L\p,M\p}}\rom{CMB}\right) 
}
Note that the above equation can be expressed without the quantity, 
$\ave{\cdots}\rom{CMB}$. For instance, in the case using the temperature 
anisotropies alone (i.e., $\alpha=\Theta\Theta$), the above equation 
is rewritten as 
\al{
  n^{x,(\Theta\Theta)}_{\ell,m,L,L\p} 
    &= (-1)^m \frac{2\ell+1}{f^{x,(\Theta\Theta)}_{\ell,L,L\p}}\sum_{M,M\p}
      \Wjm{\ell}{L}{L\p}{-m}{M}{M\p}\bigg\{
      \Theta_{L,M}\Theta_{L\p,M\p}-C_L^{\Theta\Theta}\delta_{L,L\p}\delta_{M,-M\p}(-1)^M 
    \notag \\ 
    &\qquad + 
      \sum_{x\p=\grad,\curl}\sum_{\ell\p,m\p}
      \bigg\{\sum_{L\pp,M\pp}
        \Big[
        (-1)^{M\p}\Wjm{L\p}{\ell\p}{L\pp}{-M\p}{m\p}{M\pp}\,_0\mcS_{L\p,\ell,L\pp}^{x\p}
        \Theta_{L\pp,M\pp}\Theta_{L,M}
    \notag \\ 
    &\qquad\qquad + (-1)^{M}\Wjm{L}{\ell\p}{L\pp}{-M}{m\p}{M\pp}\,_0\mcS_{L,\ell,L\pp}^{x\p}
        \Theta_{L\pp,M\pp}\Theta_{L\p,M\p}
        \Big] \notag \\ 
    &\qquad\qquad -(-1)^{m\p}\Wjm{\ell\p}{L}{L\p}{-m\p}{M}{M\p}f^{x\p,(\Theta\Theta)}_{\ell\p,L,L\p} 
      \bigg\}
      x\p_{\ell\p,m\p}
  \bigg\}
  \,.
}

The above estimator (\ref{3:est-naive}) suffers from several drawbacks 
in practical application to observation. At first, we should know about the primary 
CMB angular power spectra included in $f^{x,(\alpha)}_{\ell,L,L\p}$ a priori. 
Another problem is that the estimator includes the contribution from 
the term $n^{x,(\alpha)}_{\ell,m,L,L\p}$ which leads to a noisy 
reconstruction of the lensing potentials. Nevertheless, for the former point, 
the primary CMB angular power spectrum can be theoretically inferred if 
we know a set of fiducial cosmological parameters from other observations. 
On the other hand, for the latter point, we redefine the estimator for 
$\grad$ and $\curl$ by introducing a weight function, $F^{x,(\alpha)}_{\ell,L,L\p}$, 
in order to reduce the contribution from $n^{x,(\alpha)}_{\ell,m,L,L\p}$. 
Summing up all combination of $L$ and $L\p$, we write the estimator of 
the lensing potentials as 
\al{
  \hat{x}_{\ell,m}^{(\alpha)} 
    &= \sum_{L,L\p}F^{x,(\alpha)}_{\ell,L,L\p} \sum_{M,M\p}(-1)^m
      \Wjm{\ell}{L}{L\p}{-m}{M}{M\p}\tilde{X}_{L,M}\tilde{Y}_{L\p,M\p}  
  \,. 
  \label{3:est}  
}
The functional form of the weight function is determined so that the 
noise contribution is minimized. 

In what follows, we determine the functional form of the weight function 
so that the estimator is unbiased and the noise term is minimized. 
Eq.(\ref{3:est}) can be recast as 
\al{
  \hat{x}^{(\alpha)}_{\ell,m} 
    &= \sum_{L,L\p}F^{x,(\alpha)}_{\ell,L,L\p}\sum_{x\p=\grad,\curl}
      \frac{f^{x\p,(\alpha)}_{\ell,L,L\p}}{2\ell+1}{x\p}_{\ell,m} 
      + n^{x,(\alpha)}_{\ell,m} 
    \notag \\ 
    &= \sum_{x\p=\grad,\curl}[F^x,f^{x\p}]^{(\alpha)}_{\,\ell}x\p_{\ell,m}
      + n^{x,(\alpha)}_{\ell,m}
  \,,
  \label{3:est-expand}
}
where the inner product $[a^x,b^{x\p}]^{(\alpha)}_{\,\ell}$ for 
arbitrary two quantities, $a^{x,(\alpha)}_{\ell,L,L\p}$ and 
$b^{x\p,(\alpha)}_{\ell,L,L\p}$, is defined by 
\beq 
  [a^x,b^{x\p}]^{(\alpha)}_{\,\ell} \equiv 
    \frac{1}{2\ell+1}\sum_{L,L\p}
    a^{x,(\alpha)}_{\ell,L,L\p}b^{x\p,(\alpha)}_{\ell,L,L\p} 
  \,.
  \label{3:bracket}
\eeq 
The quantity, $n^{x,(\alpha)}_{\ell,m}$, is defined as 
\al{ 
  n^{x,(\alpha)}_{\ell,m} &\equiv \sum_{L,L\p}F^{x,(\alpha)}_{\ell,L,L\p}
    \frac{f^{x,(\alpha)}_{\ell,L,L\p}}{2\ell+1}n^{x,(\alpha)}_{\ell,m,L,L\p}
  \,.
  \label{3:noise}
} 
Eq.(\ref{3:est-expand}) implies that the estimator would be an 
unbiased estimator if we impose the following condition:
\beq 
  [F^x,f^{x\p}]^{(\alpha)}_{\,\ell} = \delta_{x,x\p}
  \,.
  \label{3:unbias}
\eeq 
Mathematically, this is equivalent to 
$\langle \hat{x}^{(\alpha)}_{\ell,m}\rangle\rom{CMB}=x_{\ell,m}$. 
Also, we wish to suppress the noise contributions, 
$n^{x,(\alpha)}_{\ell,m}$, imposing the following condition: 
\beq 
  \DD{F_{\ell,L,L\p}^{x,(\alpha)}}\ave{|n^{x,(\alpha)}_{\ell,m}|^2} = 0 
  \,.
  \label{3:mv}
\eeq 

Let us determine the functional form of the weight function under the 
conditions, (\ref{3:unbias}) and (\ref{3:mv}), with the 
Lagrange-multiplier method. 
The variance of $n^{(\alpha)}_{\ell,m}$ is given by 
\al{
  \ave{|n^{(\alpha)}_{\ell,m}|^2} 
    &= \frac{1}{2\ell+1} 
      \sum_{L,L\p} (F_{\ell,L,L\p}^{x,(\alpha)})^* \notag \\
    &\qquad \times \left(
      F_{\ell,L,L\p}^{x,(\alpha)}\tilde{\mcC}_L^{XX\p}\tilde{\mcC}_{L\p}^{YY\p}
      + F_{\ell,L\p,L}^{x,(\alpha)} (-1)^{\ell+L+L\p}
      \tilde{\mcC}_L^{XY\p}\tilde{\mcC}_{L\p}^{X\p Y}\right) 
   \,, 
  \label{3:var}
}
where the quantity, $\tilde{\mcC}_L^{XY}$, is the lensed angular power 
spectrum including the contributions from instrumental noise. 
The detailed calculation for the noise variance, 
$\langle |n^{(\alpha)}_{\ell,m}|^2 \rangle$, is presented in 
appendix \ref{appB}. Then, Eq.(\ref{3:mv}) under the constraint 
(\ref{3:unbias}) is equivalent to 
\al{ 
  \DD{F_{\ell,L,L\p}^{x,(\alpha)}}
    &\bigg\{\frac{1}{2\ell+1}\sum_{L,L\p} (F_{\ell,L,L\p}^{x,(\alpha)})^*
      \left(
        F_{\ell,L,L\p}^{x,(\alpha)}\tilde{\mcC}_L^{XX}\tilde{\mcC}_{L\p}^{YY}
        + (-1)^{\ell+L+L\p}F_{\ell,L\p,L}^{x,(\alpha)}
        \tilde{\mcC}_L^{XY}\tilde{\mcC}_{L\p}^{XY}
      \right)
    \notag \\ 
    &\qquad + \sum_{x\p=\grad,\curl}
      \lambda^x_{x\p} \left([F^{x},f^{x\p}]^{(\alpha)}_{\,\ell}-\delta_{xx\p}\right) 
    \bigg\}
    = 0 
  \,. 
  \label{LM}
} 
The quantities, $\lambda^x_{\grad}$ and $\lambda^x_{\curl}$, are the 
Lagrange multiplier whose functional form is specified below. 
Eq.(\ref{LM}) leads to 
\beq  
  (F_{\ell,L,L\p}^{x,(\alpha)})^*\tilde{\mcC}_L^{XX}\tilde{\mcC}_{L\p}^{YY}
    + (F_{\ell,L\p,L}^{x,(\alpha)})^*(-1)^{\ell+L+L\p}
      \tilde{\mcC}_L^{XY}\tilde{\mcC}_{L\p}^{XY}
    + \sum_{x\p=\grad,\curl}\lambda^x_{x\p} f^{x\p,(\alpha)}_{\ell,L,L\p} = 0 
    \,.
  \label{L-Lp} 
\eeq 
In the above, interchanging $L$ and $L\p$, we also obtain 
\beq 
  (F_{\ell,L\p,L}^{x,(\alpha)})^*\tilde{\mcC}_{L\p}^{XX}\tilde{\mcC}_L^{YY}
    +(F_{\ell,L,L\p}^{x,(\alpha)})^*(-1)^{\ell+L+L\p}\tilde{\mcC}_L^{XY}\tilde{\mcC}_{L\p}^{XY}
    + \sum_{x\p=\grad,\curl}\lambda^x_{x\p} f^{x\p,(\alpha)}_{\ell,L\p,L} 
    = 0
    \,.
  \label{Lp-L}
\eeq 
Multiplying the factors $\tilde{\mcC}_{L\p}^{XX}\tilde{\mcC}_L^{YY}$ and 
$-(-1)^{\ell+L+L\p}\tilde{\mcC}_L^{XY}\tilde{\mcC}_{L\p}^{XY}$ with 
Eq.(\ref{L-Lp}) and Eq.(\ref{Lp-L}), respectively, the sum of 
Eqs.(\ref{L-Lp}) and (\ref{Lp-L}) gives 
\beq 
  F_{\ell,L,L\p}^{x,(\alpha)}
    + \sum_{x\p=\grad,\curl}(\lambda^x_{x\p})^* g^{x\p,(\alpha)}_{\ell,L,L\p} = 0
  \,,
  \label{F-g}
\eeq 
where we define 
\al{ 
  g_{\ell,L,L\p}^{x,(\alpha)} 
    &= \frac{(f^{x,(\alpha)}_{\ell,L,L\p})^*
      \tilde{\mcC}_{L\p}^{XX}\tilde{\mcC}_L^{YY} 
      - (-1)^{\ell+L+L\p}\tilde{\mcC}_L^{XY}
      \tilde{\mcC}_{L\p}^{XY}(f^{x,(\alpha)}_{\ell,L\p,L})^*}
      {\tilde{\mcC}_L^{XX}\tilde{\mcC}_{L\p}^{YY}
      \tilde{\mcC}_{L\p}^{XX}\tilde{\mcC}_L^{YY} 
      - (\tilde{\mcC}_L^{XY}\tilde{\mcC}_{L\p}^{XY})^2} 
    \label{full:g-func}
  \,.
}
Substituting Eq.(\ref{F-g}) into Eq.(\ref{3:unbias}), we obtain 
\beq 
  -\sum_{x\pp}(\lambda^x_{x\pp})^*[g^{x\pp},f^{x\p}]^{(\alpha)}_{\ell} 
  = \delta_{xx\p} 
  \,.
  \label{gf-delta}
\eeq  
From Eq.(\ref{3:sep}), we find 
\beq 
  [g^{x\pp},f^{x\p}]^{(\alpha)}_{\,\ell} 
    = \delta_{x\pp x\p}[g^{x\p},f^{x\p}]^{(\alpha)}_{\,\ell}
  \,.
  \label{ortho}
\eeq 
Combining the above equation with Eq.(\ref{gf-delta}), we obtain the 
explicit form of the Lagrange multiplier 
\al{ 
  (\lambda^x_{x\p})^* 
    &= -\frac{\delta_{xx\p}}{[f^x,g^x]^{(\alpha)}_{\,\ell}} 
  \,. 
  \label{lambda}
} 
Then, from Eq.(\ref{F-g}), we finally obtain the expression for the weight 
function: 
\beq 
  F_{\ell,L,L\p}^{x,(\alpha)} 
    = \frac{g^{x,(\alpha)}_{\ell,L,L\p}}{[f^x,g^x]^{(\alpha)}_{\,\ell}} 
  \,.
  \label{3:F-norm} 
\eeq 
Note that, with the explicit expression (\ref{3:F-norm}), the noise 
variance, $N^{x,(\alpha)}_{\ell}$, given in Eq.(\ref{3:var}) becomes 
\al{
  N^{x,(\alpha)}_{\ell} 
    &\equiv \ave{|n_{\ell,m}^{x,(\alpha)}|^2} 
    \notag \\ 
    &= \frac{1}{2\ell+1}\frac{1}{[f^x,g^x]^{(\alpha)}_{\,\ell}} 
      \sum_{L,L\p} (g^{x,(\alpha)}_{\ell,L,L\p})^* 
    \notag \\
    &\qquad \times \left(
      F_{\ell,L,L\p}^{x,(\alpha)}\tilde{\mcC}_L^{XX}\tilde{\mcC}_{L\p}^{YY}
      + F_{\ell,L\p,L}^{x,(\alpha)} (-1)^{\ell+L+L\p}
      \tilde{\mcC}_L^{XY}\tilde{\mcC}_{L\p}^{XY}\right) 
    \notag \\
    &= \frac{1}{[f^x,g^x]^{(\alpha)}_{\,\ell}} 
  \label{3:N} 
  \,, 
} 
where we use the relations given in Eqs.(\ref{L-Lp}) and (\ref{lambda}). 
Thus, the weight function can be recast as 
\beq 
  F_{\ell,L,L\p}^{x,(\alpha)} 
    = N^{x,(\alpha)}_{\,\ell}g^{x,(\alpha)}_{\ell,L,L\p} 
  \,.
  \label{3:F} 
\eeq 

With the weight function given above, the estimators defined by 
Eq.(\ref{3:est}) become optimal, i.e., the noise contribution is minimized. 
Eq.(\ref{3:F-norm}) or Eq.(\ref{3:F}) is one of the main results in this paper. 
Note that, if the curl mode is absent, $\curl=0$, the resultant form 
of the weight function for $\grad$ exactly coincides with the one 
obtained in Ref.\cite{Okamoto:2003zw}. 
The difference appears when the angular power spectrum of the pseudo-scalar 
lensing potential, $C_{\ell}^{\curl\curl}$, included in the lensed angular 
power spectrum becomes non-vanishing. 
Note again that, in practical case, to use the estimator, 
the angular power spectrum of primary CMB anisotropies should be a priori 
known (i.e., $f^{\grad,(\alpha)}_{\ell,L,L\p}$ and 
$f^{\curl,(\alpha)}_{\ell,L,L\p}$ are given). 

\subsubsection{
Optimal combination
}

As discussed in Ref.\cite{Okamoto:2003zw}, 
the noise contribution can be further suppressed by combining multiple 
observables. Summing up the whole possible combination of temperature 
and polarization anisotropies, the optimal combination of the minimum 
variance estimators are given by 
\beq 
  \hat{x}_{\ell,m}^{(\rm c)} 
    = \sum_{\alpha}W^{x,(\alpha)}\hat{x}^{(\alpha)}_{\ell,m} 
    \qquad 
    \text{($x=\grad,\curl$)} 
  \,. 
\eeq 
The weight functions, $W^{x,(\alpha)}$, are determined so that the 
estimator satisfies the unbiased condition 
($\langle\hat{x}_{\ell,m}^{(\rm c)}\rangle\rom{CMB}=x_{\ell,m}$), 
and the variance of the noise contribution is minimum. 
The optimal combination of the minimum variance estimator is then 
determined by the same analogy as in Ref.\cite{Okamoto:2003zw}, and the 
result is 
\beq 
  \hat{x}_{\ell,m}^{(\rm c)} = N_{\ell}^{x,({\rm c})}
    \sum_{\alpha,\beta}\{(\bm{N}_{\ell}^x)^{-1}\}^{\alpha,\beta}
    \hat{x}_{\ell,m}^{(\alpha)} 
  \,, 
  \label{3:c}
\eeq 
where the variance, $N_{\ell}^{x,({\rm c})}$, is defined by 
\al{ 
  \frac{1}{N_{\ell}^{x,({\rm c})}} &= \sum_{\beta,\beta\p}
    \{(\bm{N}_{\ell}^x)^{-1}\}_{\beta\beta\p}
  \,. 
  \label{3:N-c}
} 
The component of the matrix, $\{\bm{N}_{\ell}^x\}_{\alpha,\beta}$, 
is the covariance of $n_{\ell,m}^{x,(\alpha)}$ and $n_{\ell,m}^{x,(\beta)}$ 
which is given by 
\al{
  N^{x,(\alpha,\beta)}_{\ell} &\equiv 
    \ave{(n^{x,(\alpha)}_{\ell,m})^*n_{\ell,m}^{x,(\beta)}} 
    \notag \\
    &= \frac{1}{2\ell+1} 
      \sum_{L,L\p} (F_{\ell,L,L\p}^{x,(\alpha)})^* \notag \\
    &\qquad \times \left(
      F_{\ell,L,L\p}^{x,(\beta)}\tilde{\mcC}_L^{XX\p}\tilde{\mcC}_{L\p}^{YY\p}
      + F_{\ell,L\p,L}^{x,(\beta)} (-1)^{\ell+L+L\p}
      \tilde{\mcC}_L^{XY\p}\tilde{\mcC}_{L\p}^{X\p Y}\right) 
   \,. 
  \label{3:covar}
}
The derivation of the above equation is given in appendix \ref{appB}. 

\subsection{
Flat-sky limit
} 

\btbc
\caption{Functional forms of $\ol{f}^{\grad,(XY)}_{\bl,\bL,\bL^{\prime}}$ 
and $\ol{f}^{\curl,(XY)}_{\bl,\bL,\bL^{\prime}}$ in the flat-sky case.} 
\vs{0.5}
\begin{tabular}{c|c|c} \hline 
$XY$ & $\ol{f}^{\grad,(XY)}_{\bl,\bL,\bL\p}$ & 
$\ol{f}^{\curl,(XY)}_{\bl,\bL,\bL\p}$ 
\\ \hline \\
$\Theta\Theta$ & 
$\uTT_{L}\bl\cdot\bL+\uTT_{L\p}\bl\cdot\bL\p$ &
$\uTT_{L}(\star\bl)\cdot\bL+\uTT_{L\p}(\star\bl)\cdot\bL\p$ 
\\ \\ \hline \\ 
$\Theta E$ & 
$\uTE_{L}\bl\cdot\bL\cos2\varphi_{L,L\p}+\uTE_{L\p}\bl\cdot\bL\p$ &
$\uTE_{L}(\star\bl)\cdot\bL\cos2\varphi_{L,L\p}+\uTE_{L\p}(\star\bl)\cdot\bL\p$ 
\\ \\ \hline \\ 
$\Theta B$ & 
$\uTE_{L}\bl\cdot\bL\sin 2\varphi_{L,L\p}$ & 
$\uTE_{L}(\star\bl)\cdot\bL\sin 2\varphi_{L,L\p}$ 
\\ \\ \hline \\
$EE$ & 
$[\bl\cdot\bL\uEE_{L}+\bl\cdot\bL\p\uEE_{L\p}]\cos2\varphi_{L,L\p}$ &
$[(\star\bl)\cdot\bL\uEE_{L}+(\star\bl)\cdot\bL\p\uEE_{L\p}]\cos2\varphi_{L,L\p}$ 
\\ \\ \hline \\ 
$EB$ & 
$[\bl\cdot\bL\uEE_{L}-\bl\cdot\bL\p\uBB_{L\p}]\sin2\varphi_{L,L\p}$ &
$[(\star\bl)\cdot\bL\uEE_{L}-(\star\bl)\cdot\bL\p\uBB_{L\p}]\sin2\varphi_{L,L\p}$ 
\\ \\ \hline \\ 
$BB$ & 
$[\bl\cdot\bL\uBB_{L}+\bl\cdot\bL\p\uBB_{L\p}]\cos2\varphi_{L,L\p}$ &
$[(\star\bl)\cdot\bL\uBB_{L}+(\star\bl)\cdot\bL\p\uBB_{L\p}]\cos2\varphi_{L,L\p}$ 
\\ \\ \hline 
\end{tabular}
\label{table:olf}
\etec

The quadratic estimator for the curl mode has been empirically derived in 
previous work \cite{Cooray:2005hm}, based on the flat-sky approximation. 
Here, we show that our full-sky estimator can reproduce the flat-sky 
estimator of Ref.\cite{Cooray:2005hm} (Eqs.(10)-(12) of Ref.\cite{Cooray:2005hm}), 
in the flat-sky limit, $\ell,L,L\p\ll1$.

Let us first rewrite the full-sky estimator (\ref{3:est}) in Fourier space. 
In the flat-sky limit, we usually adopt the plane wave as a harmonic basis, 
and spin-$0$ quantity, $A(\hatn)$, such as, $\Theta$, $\grad$ and $\curl$, 
is described in Fourier space as \cite{Hu:2000ee}
\al{
  A_{\bl} = \int d^2\hatn e^{-i\bl\cdot\hatn} A(\hatn) 
  \,, 
  \label{flat:spin0}
}
where the two-dimensional vector, $\bl$, is given by 
$(\ell\cos\varphi_{\ell},\ell\sin\varphi_{\ell})$. 
Similarly, in the flat-sky limit, $E$, and $B$ modes are related to the 
Stokes parameters $Q(\hatn)$ and $U(\hatn)$ as \cite{Hu:2000ee}
\al{
  E_{\bl} \pm iB_{\bl} = -\int d^2\hatn 
    e^{\pm 2i(\varphi-\varphi_{\ell})} e^{-i\bl\cdot\hatn} [Q\pm iU](\hatn) 
  \,. 
  \label{flat:spin2}
}
Eqs.(\ref{flat:spin0}) and (\ref{flat:spin2}) imply that the Fourier 
coefficients $Z_{\bl}$ ($=A_{\bl}$, $E_{\bl}$ and $B_{\bl}$) are related 
to the harmonic coefficients, $Z_{\ell,m}$ 
($=A_{\ell,m}$, $E_{\ell,m}$ and $B_{\ell,m}$), through \cite{Hu:2000ee}: 
\al{
  Z_{\bl} &= \sqrt{\frac{4\pi}{2\ell+1}}
    \sum_m i^{-m}Z_{\ell,m}e^{im\varphi_{\ell}} 
  \,, 
  \label{abl-alm} 
  \\
  Z_{\ell,m} &= \sqrt{\frac{2\ell+1}{4\pi}} i^m
    \int\frac{d\varphi_{\ell}}{2\pi} e^{-im\varphi_{\ell}}Z_{\bl} 
  \,. 
  \label{alm-abl}
}
Using the above equations (\ref{abl-alm}) and (\ref{alm-abl}), and 
combining Eq.(\ref{3:F}), the full-sky estimator (\ref{3:est}) 
in Fourier space is re-expressed as 
\beq 
  \hat{x}_{\bl}^{(\alpha)} = \sum_{L,L\p}LL\p
    \int\frac{d\varphi_L}{2\pi}\int\frac{d\varphi_{L\p}}{2\pi}
    \mcT_{\bl,\bL,\bL\p}N_{\ell}^{x,(\alpha)}g^{x,(\alpha)}_{\ell,L,L\p}
    \tilde{X}_{\bL}\tilde{Y}_{\bL\p}
  \,, 
  \label{flat:est0}
\eeq 
where we define 
\al{ 
  \mcT_{\bl,\bL,\bL\p} 
    &\equiv \left(\frac{(2L+1)(2L\p+1)}{4\pi(2\ell+1)(LL\p)^2}\right)^{1/2}
      \sum_{m,M,M\p}(-1)^m\Wjm{\ell}{L}{L\p}{-m}{M}{M\p}
      \notag \\ 
    &\qquad\qquad\qquad\qquad\qquad\qquad\qquad \times
      e^{im\varphi_{\ell}}e^{-iM\varphi_L}
      e^{-iM\p\varphi_{L\p}} i^{-m+M+M\p}
  \,. 
  \label{T-func}
} 

To go further, we approximate the quantities, 
$\mcT_{\bl,\bL,\bL\p}g^{x,(\alpha)}_{\ell,L,L\p}$ and 
$N_{\ell}^{x,(\alpha)}$, taking the flat-sky limit. 
To do this, we use the following relation valid under the flat-sky approximation, 
$\ell\gg1$ \cite{Hu:2000ee}: 
\al{
	e^{\pm si(\varphi_{\ell}-\varphi)}e^{i\bl\cdot\hatn} 
    &\simeq (\pm i)^{s}\sqrt{\frac{2\pi}{\ell}}
      \sum_m i^m\,_{\pm s}\mcY_{\ell,m}(\hatn)e^{-im\varphi_{\ell}} 
  \quad 
  (s=0,2)
  \,.
  \label{exp-Ylm} 
}
We also note that the delta function is given by \cite{Hu:2000ee}:
\beq 
  \delta_{\bl}
    = \Int{\hatn} e^{i\bl\cdot\hatn} 
  \,. 
  \label{delta} 
\eeq 
Using Eqs.(\ref{delta}), (\ref{exp-Ylm}), (\ref{mcSg}), (\ref{mcSc}), 
(\ref{2mcSg}), and (\ref{2mcSc}), we find that, under the flat-sky approximation, 
\al{
  \mcT_{\bl,\bL,\bL\p}g^{x,(\alpha)}_{\ell,L,L\p}
    &\simeq\delta_{\bL+\bL\p-\bl}\left\{\frac{\tilde{\mcC}^{XX}_{L\p}\tilde{\mcC}^{YY}_L
      (\ol{f}^{x,(\alpha)}_{\bl,\bL,\bL\p})^* 
      - \tilde{\mcC}_{L}^{XY}\tilde{\mcC}_{L\p}^{XY}
      (\ol{f}^{x,(\alpha)}_{\bl,\bL\p,\bL})^*}
    {\tilde{\mcC}^{XX}_{L\p}\tilde{\mcC}^{YY}_L
     \tilde{\mcC}^{XX}_{L}\tilde{\mcC}^{YY}_{L\p}
      - (\tilde{\mcC}^{XY}_L\tilde{\mcC}^{XY}_{L\p})^2} \right\}
    \notag \\ 
    &\equiv \delta_{\bL+\bL\p-\bl}\ol{g}^{x,(\alpha)}_{\bl,\bL,\bL\p} 
  \,,
  \label{flat:g-func} 
  \\
  N_{\ell}^{x,(\alpha)}
    &\simeq 
    \left\{\Int{\bL}\int d^2\bL\p \delta_{\bL+\bL\p-\bl} 
    \ol{f}^{x,(\alpha)}_{\bl,\bL,\bL\p}\ol{g}^{x,(\alpha)}_{\bl,\bL,\bL\p}\right\}^{-1}
    \notag \\
    &\equiv \ol{N}_{\ell}^{x,(\alpha)}
  \,, 
  \label{flat:N}
}
where the function, $\ol{f}^{x,(\alpha)}_{\bl,\bL,\bL\p}$, is given 
in Table.\ref{table:olf} for each $x$ and $\alpha$. 
Note that the quantity, $\ol{N}_{\ell}^{x,(\alpha)}$, is the flat-sky 
counterpart of the minimum variance, $N_{\ell}^{x,(\alpha)}$. 
The detailed derivation of Eqs.(\ref{flat:g-func}) and (\ref{flat:N}) 
is given in appendix \ref{appC}. 
Then, Eq.(\ref{flat:est0}) is rewritten as 
\beq 
  \hat{x}_{\bl}^{(\alpha)} = \Int{\bL}\int d^2\bL\p \delta_{\bL+\bL\p-\bl} 
    \ol{F}^{x,(\alpha)}_{\bl,\bL,\bL\p}\tilde{X}_{\bL}\tilde{Y}_{\bL\p} 
    \qquad (x=\grad,\curl) 
    \,,
    \label{flat:est}
\eeq  
where we define the function, $\ol{F}^{x,(\alpha)}_{\bl,\bL,\bL\p}$, as 
\beq 
  \ol{F}^{x,(\alpha)}_{\bl,\bL,\bL\p}
    =\ol{N}_{\ell}^{x,(\alpha)}\ol{g}^{x,(\alpha)}_{\bl,\bL,\bL\p}
  \,.
  \label{flat:F}
\eeq 
Eq.(\ref{flat:est}) is the flat-sky counterpart of Eq.(\ref{3:est}),  
which exactly coincides with that empirically defined in 
Ref.\cite{Cooray:2005hm}. Ref.\cite{Cooray:2005hm} mentioned that their 
estimator does not satisfy the unbiased condition, and may detect some 
non-zero signals of $\curl$ even in the absence of the curl mode. 
But our result show that their estimator satisfies the condition, 
$\langle \hat{x}_{\bl}^{(\alpha)}\rangle\rom{CMB}=x_{\bl}$, and becomes 
zero when the pseudo-scalar lensing potential vanishes
\footnote{In fact, the integrand in Eq.(14) of Ref.\cite{Cooray:2005hm} 
is an odd function in terms of the angle of $\bl_1$, and the right-hand 
side of Eq.(14) vanishes.}
.

Before closing this section, we also give the expression for the optimal 
combination of the flat-sky estimator used in the next section. 
With Eq.(\ref{alm-abl}), the optimal combination (\ref{3:c}) is rewritten as 
\beq 
  \hat{x}_{\bl}^{(\rm c)} = N_{\ell}^{x,({\rm c})}
    \sum_{\alpha,\beta}\{(\bm{N}_{\ell}^x)^{-1}\}^{\alpha,\beta}
    \hat{x}_{\bl}^{(\alpha)} 
  \,, 
\eeq 
and the minimum variance is obtained from Eq.(\ref{3:N-c}). 
Then, if we denote the flat-sky counterpart of $N_{\ell}^{x,({\rm c})}$ 
and $\bm{N}_{\ell}^x$ as $\ol{N}_{\ell}^{x,({\rm c})}$ 
and $\ol{\bm{N}}_{\ell}^x$, respectively, the optimal combination in the 
flat-sky limit is described by 
\beq 
  \hat{x}_{\bl}^{(\rm c)} = \ol{N}_{\ell}^{x,({\rm c})}
    \sum_{\alpha,\beta}\{(\ol{\bm{N}}_{\ell}^x)^{-1}\}^{\alpha,\beta}
    \hat{x}_{\bl}^{(\alpha)} 
  \,, 
  \label{flat:c}
\eeq 
with the variance, $\ol{N}_{\ell}^{x,({\rm c})}$, given by 
\al{ 
  \frac{1}{\ol{N}_{\ell}^{x,({\rm c})}} &= \sum_{\beta,\beta\p}
    \{(\ol{\bm{N}}_{\ell}^x)^{-1}\}_{\beta\beta\p}
  \,. 
  \label{flat:N-c}
} 
The component of the matrix, $\{\ol{\bm{N}}_{\ell}^x\}_{\alpha,\beta}$, 
is obtained by computing the flat-sky counterpart of Eq.(\ref{3:covar}) 
and the result is 
\al{ 
  \ol{N}_{\ell}^{x,(\alpha,\beta)} &= 
    \Int{L}\int d^2\bL\p \delta_{\bL+\bL\p-\bl} 
    (\ol{F}^{x,(\alpha)}_{\bl,\bL,\bL\p})^* \notag \\ 
    &\qquad\times 
    \left(\ol{F}^{x,(\beta)}_{\bl,\bL,\bL\p}\mcC_L^{XX\p}\mcC_{L\p}^{YY\p}
    +\ol{F}^{x,(\beta)}_{\bl,\bL\p,\bL}\mcC_L^{XY\p}\mcC_{L\p}^{X\p Y}\right)
  \,. 
  \label{flat:covar}
}  
The detailed calculation of Eq.(\ref{flat:covar}) is given in appendix 
\ref{appC}.

\section{
Noise spectrum
}
\label{sec.4} 

In this section, as a first step to estimate the feasibility to detect 
the curl mode based on the quadratic estimator, we compute the noise 
spectrum of the full-sky estimator, in the following cases; ACTPol 
combined with PLANCK (ACTPol+PLANCK), and cosmic-variance limit (CV-limit). 
We also numerically evaluate the difference between the full- and flat-sky 
noise spectra. 

\btbc
\caption{
\label{CMB-noise}
Experimental specifications for the PLANCK and ACTPol used in this paper. 
The quantity $\theta_\nu$ is the beam size, and $\sigma_{\nu}$ 
represents the sensitivity of each channel to the temperature 
$\sigma_{\nu,T}$ or polarizations $\sigma_{\nu,P}$, depending on the 
power spectrum of temperature ($X=\Theta$) or polarizations 
($X=E$ or $B$). The quantity $\nu$ means a channel frequency. 
}
\vs{0.5}
\begin{tabular}{cccccc} \hline 
Experiment & $f\rom{sky}$ & $\nu$ [GHz] & $\theta_{\nu}$ [arcmin] 
& $\sigma_{\nu,T}$ [$\mu$K/pixel] & $\sigma_{\nu,P}$ [$\mu$K/pixel] 
\\ \hline 
\PL \cite{PLANCK} & 0.65 & 30 & 33 & 4.4 & 6.2 \\ 
 & & 44 & 23 & 6.5 & 9.2  \\ 
 & & 70 & 14 & 9.8 & 13.9 \\
 & & 100 & 9.5 & 6.8 & 10.9 \\
 & & 143 & 7.1 & 6.0 & 11.4 \\
 & & 217 & 5.0 & 13.1 & 26.7 \\
 & & 353 & 5.0 & 40.1 & 81.2 \\ \hline 
\AP \cite{ACTPol} & 0.1 & 148 & 1.4 & 3.6 & 5.0 \\ \hline 
\end{tabular}
\etec

\bfbc
\includegraphics[width=120mm,clip]{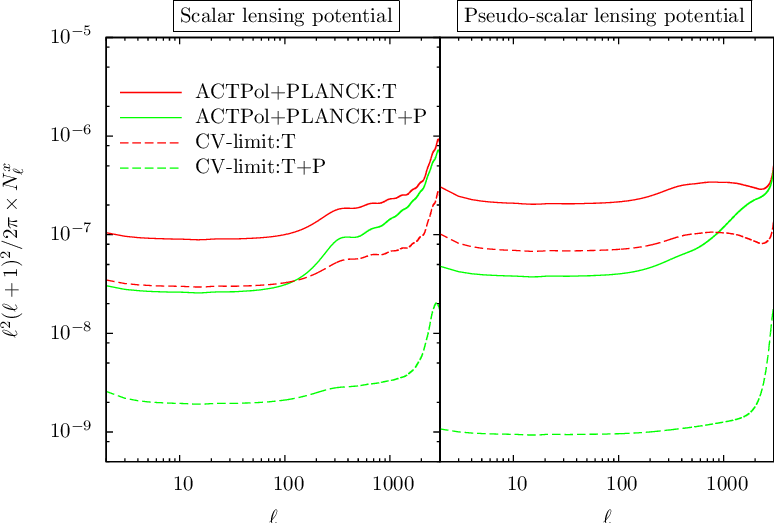}
\caption{
The noise spectra of the scalar (left) and pseudo-scalar (right) lensing 
potentials for the lensing reconstruction from the temperature map alone, 
$\alpha=\Theta\Theta$ 
(T; red lines), and from the temperature and polarization maps 
(T+P; green lines). We assume two cases; ACTPol combined with PLANCK 
(ACTPol+PLANCK; solid lines), and cosmic-variance limit (CV-limit; dashed lines). 
We compute the noise spectra according to Eqs.(\ref{3:N}) for 
T and Eq.(\ref{3:N-c}) for T+P, with $\ell\rom{max}=3000$, 
and take into account the effect of finite sky coverage. 
}
\label{Noise}
\efec

\bfbc
\includegraphics[width=120mm,clip]{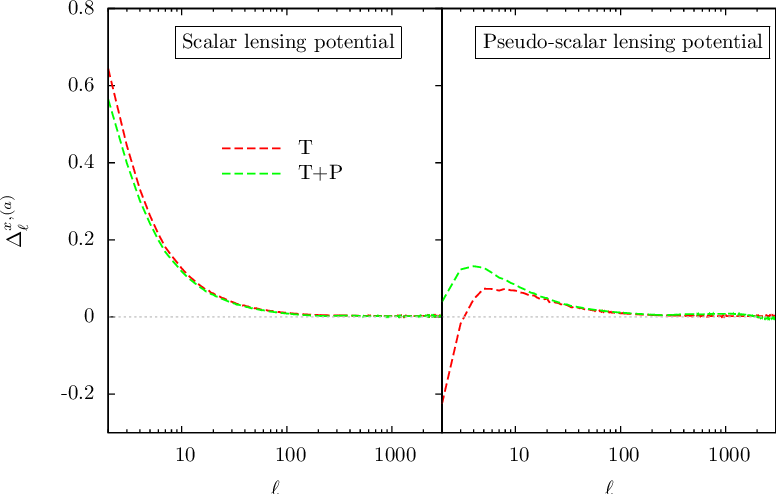}
\caption{
The fractional difference of the noise spectrum between full- and 
flat-sky estimators (\ref{frac-dif}), in the case with ACTPol+PLANCK. 
The left and right panels show the fractional difference for scalar 
and pseudo-scalar lensing potentials, respectively. 
}
\label{deltan}
\efec

In the full-sky case, the noise spectrum is given by Eqs.(\ref{3:N}) and 
(\ref{3:N-c}). On the other hand, the noise in the flat-sky limit is 
obtained from Eqs.(\ref{flat:N}) and (\ref{flat:N-c}). To compute the 
noise spectrum, we assume that the lensing effect on the CMB comes only 
from the large-scale structure, and no source to produce the 
pseudo-scalar lensing potential is present. We use the modified version 
of FuturCMB code \footnote{http://lpsc.in2p3.fr/perotto/} 
\cite{Perotto:2006rj}. In computing noise spectra, we further need the 
lensed, unlensed and instrumental noise angular power spectra. The lensed 
and unlensed power spectra are computed by CAMB \cite{Lewis:1999bs} with 
the fiducial value of cosmological parameters described in section 
\ref{sec.1}. Owing to the assumption, the angular power spectrum for the 
pseudo-scalar lensing potential, $C_{\ell}^{\varpi\varpi}$, is set to 
zero. On the other hand, the angular power spectrum of the scalar lensing 
potential is obtained by integrating the matter power spectrum along the 
line-of-site, for which we adopt the fitting formula of the non-linear 
matter power spectrum given in Ref.\cite{Smith03}. The instrumental noise 
power spectra are given by \cite{Knox:1995dq} 
\al{
  \mcN_{\ell}^{XX} 
    &= \left[\sum_{\nu}(\mcN_{\ell,\nu}^{XX})^{-1}\right]^{-1}; 
    & \mcN_{\ell,\nu}^{XX} 
    & \equiv \left(\frac{\sigma_{\nu}\theta_{\nu}}{T\rom{CMB}}\right)^2
    \exp\left[\frac{\ell(\ell+1)\theta_{\nu}^2}{8\ln 2}\right] 
  \,, 
  \label{noise}
}
with $T\rom{CMB}=2.7$K being mean temperature of CMB. Here, the quantity 
$\theta_\nu$ is the beam size, and $\sigma_{\nu}$ represents the 
sensitivity of each channel to the temperature $\sigma_{\nu,T}$ or 
polarizations $\sigma_{\nu,P}$. The specific values for PLANCK and ACTPol 
are summarized in Table \ref{CMB-noise}. 
Note that, for ACTPol+PLANCK, we assume that the survey region of ACTPol 
is entirely overlapped with that of PLANCK, and plot the following noise 
spectrum 
\al{
  N_{\ell}^{x,(a)} \equiv 
    \left[\frac{f\rom{sky}^{\rm ACTPol}}{(N^{x,(a)}_{\ell:{\rm ACTPol}})^2}
    +\frac{(f\rom{sky}^{\rm PLANCK}-f\rom{sky}^{\rm ACTPol})}
          {(N^{x,(a)}_{\ell:{\rm PLANCK}})^2}\right]^{-1/2} 
  \,,
}
where $f^{\rm ACTPol}\rom{sky}=0.1$ and $f^{\rm PLANCK}\rom{sky}=0.65$ are 
the fractional sky-coverage of ACTPol and PLANCK, respectively, and the 
label, $a$, represents a pair of temperature maps, ``$\Theta\Theta$'' or 
the optimal combination, ``c''. 
In computing the noise of reconstruction within the survey region of 
ACTPol, $N^{x,(a)}_{\ell:{\rm ACTPol}}$, we use the lensed angular power 
spectra from PLANCK instead of ACTPol, at $\ell<700$, in order to remedy a 
large uncertainty at large angular scales arising from the atmospheric 
temperature fluctuations. On the other hand, the noise, 
$N^{x,(a)}_{\ell:{\rm PLANCK}}$, is calculated with PLANCK experimental 
specification. 
For cosmic-variance limit, the reconstruction noise is computed with the 
instrumental noise being $\mcN_{\ell}^{XX}=0$. 

In Fig.\ref{Noise}, we plot the expected noise spectrum in the full-sky 
case. 
The left and right panels show the noise spectra, $N^{\grad,(a)}_{\ell}$ 
and $N^{\curl,(a)}_{\ell}$, respectively. 
The resultant noise spectra for the pseudo-scalar lensing potential have 
amplitude comparable to those for the scalar lensing potential. 
In the cosmic-variance limit, the reconstruction noise is improved 
by more than an order of magnitude compared to the case with 
ACTPol+PLANCK.
Notice that the noise spectra for the pseudo-scalar lensing potential is 
sensitive to the inclusion of polarizations compared to the estimator of 
the scalar lensing potential. 
In our calculation, the angular power spectrum of the pseudo-scalar 
lensing potential is set to zero. 
Although the primordial gravitational-waves or cosmic strings induces the 
pseudo-scalar lensing potential, as shown in the next section, the 
amplitude of $C^{\curl\curl}_{\ell}$ is at least two orders of magnitude 
smaller than that of $C^{\grad\grad}_{\ell}$, as long as we consider the 
model parameters consistent with observations. 
Thus, the inclusion of the pseudo-scalar lensing potential would hardly 
change the result in Fig.\ref{Noise}. 

In Fig.\ref{deltan}, to show the difference of the noise spectra between 
the flat- and full-sky estimators, we consider ACTPol+PLANCK and plot the 
following quantity as a function of $\ell$: 
\beq 
  \Delta^{x,(a)}_{\ell} \equiv 
    \frac{\ol{N}_{\ell}^{x,(a)}-N_{\ell}^{x,(a)}}
    {N_{\ell}^{x,(a)}} 
  \,.
  \label{frac-dif}
\eeq 
The fractional difference becomes significant at large scales, and the 
difference becomes $\gsim10$\% at $\ell\lsim10$ for both $\grad$ and 
$\curl$. 
The results do not sensitively depend on whether we include the 
polarization data for the reconstruction or not. 
Note that the fractional difference for $\grad$ is roughly consistent 
with Ref.\cite{Okamoto:2003zw}. 
These results show that the flat-sky estimator becomes invalid on large 
scales ($\ell\lsim10$) and the full-sky lensing reconstruction is essential 
to extract information on primordial gravitational-waves and cosmic strings 
as discussed in the next section.

\section{Implications for primordial gravitational-waves and cosmic strings} \label{sec.5} 

In this section, we illustrate the usefulness of the pseudo-scalar 
lensing potential as a diagnosis of the vector/tensor perturbations. 
Here we specifically focus on the pseudo-scalar lensing potential 
induced by two cases; primordial gravitational-waves produced during 
inflation, and cosmic strings (e.g., Refs.\cite{Yamauchi:2010ms,Yamauchi:2010vy}). 

\bfbc
\includegraphics[width=140mm,clip]{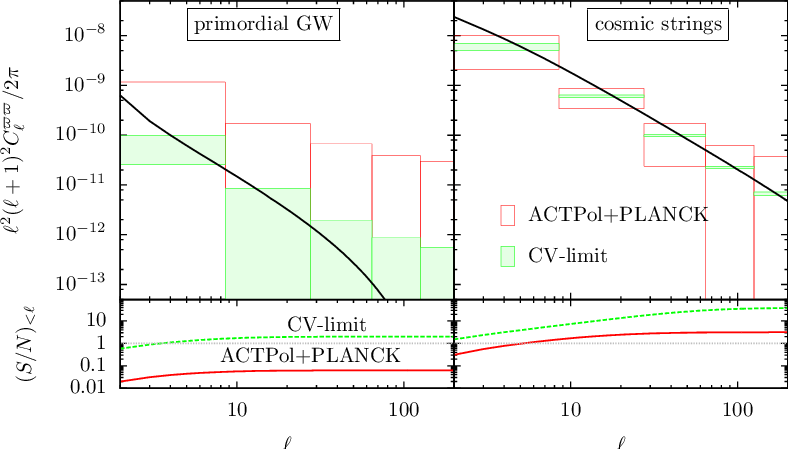}
\caption{
The angular power spectrum of the pseudo-scalar lensing potential from 
primordial gravitational-waves with the tensor-to-scalar ratio, $r=0.1$ 
(top left), and cosmic strings with $G\mu=10^{-8}$ and $P=0.001$ (top right). 
The error boxes in each figure show the expected variance 
of angular power spectrum from ACTPol combined with PLANCK (red) and 
cosmic-variance limit (green). The bottom two panels show the 
signal-to-noise ratio as a function of maximum multipole, for primordial 
gravitational-waves (bottom left) and cosmic strings (green). 
}
\label{SNR}
\efec

The gravitational waves can produce the metric perturbations which have odd 
parity symmetry. This means that the lensing effect induced by the 
gravitational waves cause the curl mode of deflection angle. 
There are several studies on the curl mode of deflections 
(or the pseudo-scalar lensing potential) induced by the primordial 
gravitational-waves 
(\cite{Dodelson:2003bv,Sarkar:2008ii,Cooray:2005hm,Li:2006si}), 
and the angular power spectrum of the pseudo-scalar lensing potential 
is given by \cite{Sarkar:2008ii}
\beq 
  C_{\ell}^{\varpi\varpi} 
    = \frac{8}{\pi\ell^2(\ell+1)^2}\frac{(\ell+2)!}{(\ell-2)!} rA_{\rm s}
    \int \frac{dk}{k}[\Delta^{\rm GW}_{\ell}(k,\eta)]^2 
  \,,
\eeq 
with 
\beq 
  \Delta^{\rm GW}_{\ell}(k,\eta) 
    = \int_{\eta_*}^{\eta_0} k d\eta \left(\frac{3j_1(k\eta)}{k\eta}\right)
    \frac{j_{\ell}(x)}{x^2}\bigg|_{x=k(\eta_0-\eta)} 
  \,.
\eeq 
The quantity, $\eta_*$, denotes the conformal time at the last scattering 
surface, and the quantity $r$ is 
the tensor-to-scalar ratio and we assume $r=0.1$ which is close to the 
current upper bound \cite{Komatsu:2010fb}. 

On the other hand, extended objects such as cosmic strings produce not only 
the scalar perturbations but also the vector/tensor perturbations, and 
induce the pseudo-scalar lensing potential. 
For simplicity, we focus on the pseudo-scalar lensing potential only from 
vector perturbations. 
To compute the pseudo-scalar lensing potential, we consider a string 
network described by the velocity-dependent one-scale model 
\cite{Martins:1996jp,Martins:2000cs,Avgoustidis:2005nv,Takahashi:2008ui,
Yamauchi:2010ms,Yamauchi:2010vy}, characterized by the intercommuting 
probability, $P$. 
The loop formation process is required for the energy loss of cosmic 
strings and the scaling solution. 
Also, this process is related to the number density of strings. 
Note that the constraint on $P$ has an important role to distinguish 
between the cosmic strings as the topological defect ($P=1$) and cosmic 
super-strings generated from the stringy inflation ($P\ll1$) 
\cite{Copeland:2003bj,Jackson:2004zg}. 
We assume that the string segments are distributed randomly between the 
last scattering surface and observer, consistently with the scaling model, 
or, in other word, we neglect the correlation between two different 
segments \cite{Yamauchi:2010vy}.

If the string tension, $G\mu$, and intercommuting probability, $P$, are 
given, the pseudo-scalar lensing potential from cosmic strings are 
computed as follows \cite{Yamauchi:2011}: 
\beq 
  C_{\ell}^{\varpi\varpi} 
    = \frac{4}{\ell(\ell+1)}(16\pi G\mu)^2\sqrt{\frac{2}{3\pi}}\frac{v^2}{1-v^2}
    \int \frac{dk}{k} [\Delta^{\rm CS}_{\ell}(k,\eta)]^2 
  \,,
\eeq 
with \cite{Yamauchi:2011}
\beq 
  \Delta^{\rm CS}_{\ell}(k,\eta) 
    = \int_{\eta_*}^{\eta_0} k d\eta 
      \left[\frac{4\pi a^4 k}{H}\left(\frac{a}{k\xi}\right)^5{\rm erf}
      \left(\frac{k\xi}{2\sqrt{6}\,a}\right)\right]^{1/2}
      j_{\ell}(x)\bigg|_{x=k(\eta_0-\eta)} 
  \,. 
\eeq 
The quantity, $H$, is the Hubble expansion rate, and the quantities, $\xi$ 
and $v$, are the correlation length and root-mean-square velocity, 
respectively, and determined from \cite{Yamauchi:2010ms} 
\al{ 
  &2v^2\left(1+\frac{\pi}{2\sqrt{2}}0.23P\frac{1+8v^6}{1-8v^6}\right)=1 
  \,,
  \label{vrms} 
  \\
  &\xi = \frac{2\sqrt{2}}{\pi}\frac{1-8v^6}{2Hv(1+8v^6)} 
  \,. 
  \label{xi}
} 
The detailed description of model for the pseudo-scalar lensing potential 
from cosmic strings is described in the forthcoming paper 
\cite{Yamauchi:2011}.

In the left two panels of Fig.\ref{SNR}, the angular power spectrum from 
primordial gravitational-waves and the signal-to-noise ratio are shown. 
The plotted errors are estimated from 
\beq
  \Delta C_{\ell} = 
    \frac{C_{\ell}^{\curl\curl}+N_{\ell}^{\curl,({\rm c})}}
    {\sqrt{(\ell+1/2)f\rom{sky}\Delta\ell}} 
  \,,
  \label{DeltaC}
\eeq
where $\Delta\ell$ is the size of multipole bin, and we set 
$\Delta\ell=(i+1)^3-i^3$ for $i$-th bin, just for illustration. 
For ACTPol combined with PLANCK, we evaluate the errors as 
\beq 
  \Delta C_{\ell} 
    = \{(\Delta C_{\ell:{\rm ACTPol}})^{-2}+(\Delta C_{\ell:{\rm PLANCK}})^{-2}\}^{-1/2}
  \,, 
\eeq 
where the errors arising from ACTPol survey region, 
$\Delta C_{\ell:{\rm ACTPol}}^{XY}$, are computed according to 
Eq.(\ref{DeltaC}) with $f\rom{sky}=0.1$ and 
$N_{\ell}^{\curl}=N^{\curl}_{\ell:{\rm ACTPol}}$.
Similarly, the errors from PLANCK survey area, $\Delta C_{\ell:{\rm PLANCK}}^{XY}$, 
are obtained from Eq.(\ref{DeltaC}) with $f\rom{sky}=0.55$ and 
$N_{\ell}^{\curl}=N^{\curl}_{\ell:{\rm PLANCK}}$. 
On the other hand, for the cosmic-variance limit, we compute the errors 
with $f\rom{sky}=1.0$ and the instrumental noise power spectra being zero. 
The signal-to-noise ratio for angular power spectrum of 
the pseudo-scalar lensing potential is defined by 
\beq 
  \left({\rm\frac{S}{N}}\right)_{<\ell}=\left\{\sum_{\ell\p=2}^{\ell}
    \left(\frac{C_{\ell\p}^{\curl\curl}}{\Delta C_{\ell\p}}\right)^2\right\}^{1/2} 
  \,.
\eeq 

As is expected, it is hard to detect the signature of primordial 
gravitational-waves from lensing reconstruction. For ACTPol+PLANCK, the 
signal-to-noise ratio is less than $0.1$. Even with CV-limit, the 
signal-to-noise ratio is $\sim 2$. For the tensor-to-scalar ratio below 
the current upper limit \cite{Komatsu:2010fb}, the signal-to-noise ratio 
would be further worsen. This is true as long as we adopt the quadratic 
estimator. These conclusions are consistent with the one obtained in 
Ref.\cite{Cooray:2005hm}, where they discussed the detectability in the 
flat-sky limit 
\footnote{
Note that there was a discrepancy 
between the results of Ref.\cite{Book:2011na} and our previous version. 
After discussion with the author of Ref.\cite{Book:2011na}, 
however, we found that the discrepancy 
of the results comes from their miscalculation and the conclusion obtained 
in our previous version is unchanged. 
}.

On the other hand, in the right panels of Fig.\ref{SNR}, we show the case 
of the cosmic strings, with $G\mu=10^{-8}$ and $P=0.001$ 
\cite{Yamauchi:2011cu}. Note that these values are consistent with the 
constraint from the temperature angular power spectrum using the 
Gott-Kaiser-Stebbins (GKS) effect \cite{Kaiser:1984,Gott:1985} induced by 
the gravitational potential of a moving string \cite{Yamauchi:2010ms}. 
Although the result depends on the parameters of comic strings, the 
signal-to-noise ratio is $\sim 3$ even for ACTPol+PLANCK case. In the 
cosmic variance limit, the signal-to-noise ratio becomes $\sim 30$. Note 
that, we have ignored the tensor metric perturbations from cosmic strings 
which also induce the pseudo-scalar lensing potential, and the 
inclusion of the contributions from the tensor perturbations would further 
increase the signal-to-noise ratio. The GKS effect 
observed via temperature map would have larger signal-to-noise 
ratio than the pseudo-scalar lensing potential \cite{Yamauchi:2010ms}. 
However, the temperature anisotropies at small angular scales are usually 
dominated by the contributions from point sources and the 
Sunyaev-Zel'dovich (SZ) effect \cite{Zeldovich:1969}. In this respect, the 
reconstruction of pseudo-scalar lensing potential is useful to check 
systematics and biases in the derived constraints on cosmic strings from 
GKS effect. Note finally that the information on large scales would be 
important for detecting the pseudo-scalar lensing potential from cosmic 
strings, and the full-sky formalism for lensing reconstruction is 
indispensable.

\section{
Summary and discussion
}
\label{sec.6}

In this paper, we presented a full-sky algorithm for reconstructing the 
lensing potential of scalar (gradient mode) and pseudo-scalar (curl mode) 
components. We defined the estimator as a quadratic combination of observed 
anisotropies, and introduced the weight function to reduce the noise contribution 
(see Eq.(\ref{3:est})). The resultant form of the weight function 
which minimizes the noise contribution is given by Eq.(\ref{3:F}) 
with Eqs.(\ref{3:N}), and (\ref{full:g-func}). 
Thanks to the different parity symmetry between scalar and pseudo-scalar 
lensing potentials, the gradient and curl modes can be separately reconstructed. 
Note that the quantities used to reconstruct the lensing potentials are summarized in Table.\ref{notation}. 
In the flat-sky limit, we showed that the estimator reduces to the one empirically defined in 
Ref.\cite{Cooray:2005hm}. We explicitly evaluated the noise spectra, 
and showed that the noise contribution for the pseudo-scalar lensing 
potential is comparable to that for the gradient mode. 
Further, prospects for reconstructing the curl mode is discussed, and 
signal-to-noise ratio for the pseudo-scalar lensing potential is computed, 
especially focusing on primordial gravitational-waves and cosmic strings.

In this paper, we specifically focused on the lensing reconstruction 
based on the quadratic estimator proposed in Ref.\cite{Okamoto:2003zw}. 
On the other hand, in Ref.\cite{Hirata:2003ka}, for experiments 
sensitive to $B$-mode polarization, it would be possible to 
improve the precision of the lensing potential with 
the estimator based on the likelihood analysis. 
This may be also true for the pseudo-scalar lensing potential. 
Extending the estimator of Ref.\cite{Hirata:2003ka} to the full-sky case, 
the signal-to-noise ratio estimated in section \ref{sec.5} would be improved, 
and upcoming or next generation experiments may detect the pseudo-scalar 
lensing potential even from primordial gravitational-waves. 
This will be investigated in our future work. 

Throughout this paper, we have assumed several idealizations, i.e., 
the higher-oder terms of the deflection angle are negligible, and 
observed CMB maps are given on the full sky without foregrounds 
and the inhomogeneous noise. 
However, at small scales, the lowest-order 
approximation (e.g., ignoring the higher-order terms in Eq.(\ref{remap}) 
for the temperature anisotropies) 
would not be valid for high resolution experiments \cite{Amblard:2004ih}. 
The higher-order terms of deflection angle produce additional contributions 
in the lensed anisotropies, and the estimated scalar and pseudo-scalar 
lensing potential include the contributions from the higher-order terms 
of deflection angle \cite{Hanson:2010rp}. 
Also, the masking effect \cite{Bucher:2010iv,Carvalho:2010rz,Carvalho:2011gx}, 
foreground contaminations from point sources and thermal/kinematic SZ effect 
\cite{Amblard:2004ih,Das:2011ak}, and the inhomogeneous noise \cite{Hanson:2009dr} 
induce the additional non-zero off-diagonal terms in Eq.(\ref{3:XY}) and the estimated 
lensing potentials would be biased. Thus, in practical cases, 
to reduce the systematic bias in the estimated lensing potentials, 
an accurate treatment of these practical problems for the lensing 
reconstruction is required and worth investigating.


\acknowledgments 
We would like to thank Sudeep Das for helpful discussion and comments. This research was 
supported in part by the Japan Society for the Promotion of Science (JSPS) Core-to-Core Program 
``International Research Network for Dark Energy'', and Grant-in-Aid for Scientific Research on 
Priority Areas No. 467 ``Probing the Dark Energy through an Extremely Wide and Deep Survey 
with Subaru Telescope''. TN is supported in part by a grant from the Hayakawa Satio Fund awarded 
by the Astronomical Society of Japan. AT is supported in part by a Grants-in-Aid for Scientific 
Research from JSPS (No. 21740168). 

\appendix
\section{Useful formula} \label{appA}

In this appendix, following Refs.\cite{QTAM} and \cite{Hu:1997hp}, 
we summarize the formulas used in this paper. 

\subsection{Symmetry of Wigner-$3j$ symbols}

Symmetric properties of the Wigner-$3j$ symbols are described by 
\al{
	&\Wjm{\ell_1}{\ell_2}{\ell_3}{m_1}{m_2}{m_3} 
		= \Wjm{\ell_2}{\ell_3}{\ell_1}{m_2}{m_3}{m_1}
		= \Wjm{\ell_3}{\ell_1}{\ell_2}{m_3}{m_1}{m_2}
		= (-1)^{\ell_1+\ell_2+\ell_3} \Wjm{\ell_3}{\ell_2}{\ell_1}{m_3}{m_2}{m_1} 
	\,, \label{sym0} 
	\\ 
	&\Wjm{\ell_1}{\ell_2}{\ell_3}{m_1}{m_2}{m_3}
		= (-1)^{\ell_1+\ell_2+\ell_3} \Wjm{\ell_1}{\ell_2}{\ell_3}{-m_1}{-m_2}{-m_3} 
	\,. \label{sym} 
}

\subsection{Summation of Wigner-$3j$ symbols}

Throughout the paper, we frequently use the following property of the Wigner-$3j$ symbols: 
\al{ 
	&\sum_M (-1)^{L+M} \Wjm{\ell}{L}{L}{-m}{M}{-M} 
		= \delta_{\ell,0}\delta_{m,0}\sqrt{\frac{2L+1}{2\ell+1}} 
	\,, \label{Wigner:sum1} 
	\\ 
	&\sum_{M,M'} \Wjm{\ell}{L}{L'}{-m}{M}{M'}\Wjm{\ell'}{L}{L'}{-m'}{M}{M'} 
		= \frac{1}{2\ell+1}\delta_{\ell,\ell'}\delta_{m,m'} 
	\,. \label{Wigner:sum2} 
}

\subsection{Relations between Wigner-$3j$ symbols and spherical harmonics}

Let us define 
\al{
	\,_{s}\mcI^{m,m',M}_{\ell,\ell',L} 
		&= \int d\hatn \,_{s}\mcY_{L,M}^*(\hatn)
		[\bm{\nabla}\,_0\mcY_{\ell,m}(\hatn)] \cdot [\bm{\nabla}\,_{s}\mcY_{\ell',m'}(\hatn)] 
	\,, \\ 
	\,_{s}\mcJ^{m,m',M}_{\ell,\ell',L} 
		&= \int d\hatn \,_{s}\mcY_{L,M}^*(\hatn)
		[(\star\bm{\nabla})\,_0\mcY_{\ell,m}(\hatn)] \cdot [\bm{\nabla}\,_{s}\mcY_{\ell',m'}(\hatn)] 
	\,. 
}
These two integrals are related to the Wigner-$3j$ symbols as 
\al{ 
  \,_{s}\mcI^{m,m',M}_{\ell,\ell',L} 
    &= \sqrt{\frac{(2L+1)(2\ell+1)(2\ell'+1)}{16\pi}}
    \Wjm{L}{\ell}{\ell'}{-M}{m}{m'}
    \Wjm{L}{\ell}{\ell'}{s}{0}{-s}  \notag \\
    &\qquad \times [-L(L+1)+\ell(\ell+1)+\ell'(\ell'+1)] 
  \,, 
  \label{Wigner:harmonics1} \\
  \,_{s}\mcJ^{m,m',M}_{\ell,\ell',L} 
    &= -i\sqrt{\frac{(2L+1)(2\ell+1)(2\ell'+1)}{16\pi}} 
      \Wjm{L}{\ell}{\ell'}{-M}{m}{m'} \notag \\
    &\qquad \times \bigg[\sqrt{\ell(\ell+1)(\ell'+s)(\ell'-s-1)}
    \Wjm{L}{\ell}{\ell'}{s}{-1}{-s+1} 
    \notag \\
    &\qquad - \sqrt{\ell(\ell+1)(\ell'-s)(\ell'+s+1)}
      \Wjm{L}{\ell}{\ell'}{s}{1}{-s-1}\bigg] 
  \,.
  \label{Wigner:harmonics2} 
} 

\section{
Noise covariance
} 
\label{appB}

Here we derive Eqs.(\ref{3:var}) and (\ref{3:covar}). 
Since Eq.(\ref{3:var}) is obtained by setting $\beta=\alpha$, in what follows, 
we derive the expression for the noise covariance given in Eq.(\ref{3:covar}). 

\subsection{
Relation between noise and estimator covariance
}

First, we rewrite the noise variance using the variance of the estimator. 
Since $n_{\ell,m}^{x,(\alpha)}=\hat{x}_{\ell,m}^{(\alpha)}-x_{\ell,m}$, 
the noise variance is rewritten as 
\beq 
  \ave{(n_{\ell,m}^{x,(\alpha)})^*n_{\ell,m}^{x,(\beta)}}
    = \ave{(\hat{x}_{\ell,m}^{(\alpha)})^*\hat{x}_{\ell,m}^{(\beta)}}
      - \ave{(x_{\ell,m})^*\hat{x}_{\ell,m}^{(\beta)}}
      - \ave{(\hat{x}_{\ell,m}^{(\alpha)})^*x_{\ell,m}}
      + C_{\ell}^{xx}
  \,. 
  \label{appB:n-est}
\eeq 
Note here that 
\al{
  &\ave{(x_{\ell,m})^*\tilde{X}_{L,M}\tilde{Y}_{L\p,M\p}} 
    \notag \\ 
  &\quad= \sum_{x\p}\sum_{\ell\p,m\p}\sum_{L\pp,M\pp}\sum_{Z=\Theta,E,B} 
    \notag \\ 
  &\quad\qquad\times\bigg\{
    (-1)^{M\p}\Wjm{L\p}{\ell\p}{L\pp}{-M\p}{m\p}{M\pp}s_{L\p,\ell\p,L\pp}^{x\p,(ZY)}
    \ave{(x_{\ell,m})^*Z_{L\pp,M\pp}x\p_{\ell\p,m\p}X_{L,M}}
    \notag \\ 
  &\qquad\qquad + (-1)^{M}\Wjm{L}{\ell\p}{L\pp}{-M}{m\p}{M\pp}s_{L,\ell\p,L\pp}^{x\p,(ZX)}
    \ave{(x_{\ell,m})^*Z_{L\pp,M\pp}x\p_{\ell\p,m\p}Y_{L\p,M\p}}
  \bigg\}
  \,, 
  \label{appB:xXY}
}
where we use Eqs.(\ref{lens:T}), (\ref{lens:E}) and (\ref{lens:B}), 
and introduce the coefficients, $s_{L,\ell,L\p}^{x,(XY)}$, defined by 
\beq 
  s_{L,\ell,L\p}^{x,(\Theta\Theta)}=\,_0\mcS_{L,\ell,L\p}^x \,, \quad 
  s_{L,\ell,L\p}^{x,(EE)}=s_{L,\ell,L\p}^{x,(BB)}=\,_{\oplus}\mcS_{L,\ell,L\p}^x \,, \quad 
  s_{L,\ell,L\p}^{x,(EB)}=-s_{L,\ell,L\p}^{x,(BE)}=\,_{\ominus}\mcS_{L,\ell,L\p}^x \,, \quad 
\eeq 
and $s_{L,\ell,L\p}^{x,(XY)}=0$ for the other combinations of $XY$. 
Assuming that the lensing potentials and primary CMB anisotropies are a 
random Gaussian field, and the correlations between the lensing 
potentials and primary CMB anisotropies are negligible, the above 
equation (\ref{appB:xXY}) reduces to 
\al{
  \ave{(x_{\ell,m})^*\tilde{X}_{L,M}\tilde{Y}_{L\p,M\p}}
    = (-1)^m \Wjm{\ell}{L}{L\p}{-m}{M}{M\p}f_{\ell,L,L\p}^{x,(XY)}C_{\ell}^{xx}
  \,.
}
With Eq.(\ref{Wigner:sum2}), this leads to the following equation: 
\al{ 
  \ave{(x_{\ell,m})^*\hat{x}_{\ell,m}^{(\alpha)}} 
    &= \sum_{L,L\p}F^{x,(\alpha)}_{\ell,L,L\p}
      \sum_{M,M\p}\Wjm{\ell}{L}{L\p}{-m}{M}{M\p}\Wjm{\ell}{L}{L\p}{-m}{M}{M\p}
      f_{\ell,L,L\p}^{x,(XY)}C_{\ell}^{xx}
    = C_{\ell}^{xx}
  \,.
} 
As a result, Eq.(\ref{appB:n-est}) becomes 
\beq 
  \ave{(n_{\ell,m}^{x,(\alpha)})^*n_{\ell,m}^{x,(\beta)}}
    = \ave{(\hat{x}_{\ell,m}^{(\alpha)})^*\hat{x}_{\ell,m}^{(\beta)}}
      - C_{\ell}^{xx}
  \label{appB:n-cov}
  \,. 
\eeq 

\subsection{
Estimator covariance
}

Next we compute the estimator covariance, 
$\ave{(\hat{x}_{\ell,m}^{(\alpha)})^*\hat{x}_{\ell,m}^{(\beta)}}$. 
From Eq.(\ref{3:est}), the covariance is given by 
\al{
  \ave{(\hat{x}^{(\alpha)}_{\ell,m})^*\hat{x}_{\ell,m}^{(\beta)}} 
    &= \sum_{L_1,L\p_1}\sum_{L_2,L\p_2}\sum_{M_1,M_1\p}\sum_{M_2,M\p_2} 
      \Wjm{\ell}{L_1}{L\p_1}{-m}{M_1}{M\p_1} 
      \Wjm{\ell}{L_2}{L\p_2}{-m}{M_2}{M\p_2} 
    \notag \\ 
    &\qquad \times 
      (F_{\ell,L_1,L\p_1}^{x,(\alpha)})^*F_{\ell,L_2,L\p_2}^{x,(\beta)} 
      \ave{\tilde{X}_{L_1,M_1}^*\tilde{Y}_{L\p_1,M\p_1}^*
      \tilde{Z}_{L_2,M_2}\tilde{W}_{L\p_2,M\p_2}}
  \,.
  \label{appB:cov}
}
To compute the right-hand side of Eq.(\ref{appB:cov}), we need to 
evaluate the four-point correlation of the observed CMB anisotropies: 
\beq 
  \ave{\tilde{X}_{L_1,M_1}^*\tilde{Y}_{L\p_1,M\p_1}^*
  \tilde{Z}_{L_2,M_2}\tilde{W}_{L\p_2,M\p_2}}
  \,. 
  \label{4point}
\eeq 
In general, the four-point correlation (\ref{4point}) is described by 
\al{ 
  &\ave{\tilde{X}_{L_1,M_1}^*\tilde{Y}_{L\p_1,M\p_1}^*
    \tilde{Z}_{L_2,M_2}\tilde{W}_{L\p_2,M\p_2}} \notag \\
  &\qquad = \ave{\tilde{X}_{L_1,M_1}^*\tilde{Y}_{L\p_1,M\p_1}^*
    \tilde{Z}_{L_2,M_2}\tilde{W}_{L\p_2,M\p_2}}\rom{G} 
    + \ave{\tilde{X}_{L_1,M_1}^*\tilde{Y}_{L\p_1,M\p_1}^*
    \tilde{Z}_{L_2,M_2}\tilde{W}_{L\p_2,M\p_2}}\rom{C} 
  \,. 
  \label{GC}
} 
The first term is defined by 
\al{ 
  &\ave{\tilde{X}_{L_1,M_1}^*\tilde{Y}_{L\p_1,M\p_1}^*
    \tilde{Z}_{L_2,M_2}\tilde{W}_{L\p_2,M\p_2}}\rom{G} \notag \\
  &\qquad = \tilde{\mcC}^{XY}_{L_1}\tilde{\mcC}^{ZW}_{L_1\p}\Delta_{(1,1\p),(2,2\p)}
    + \tilde{\mcC}^{XZ}_{L_1}\tilde{\mcC}^{YW}_{L_1\p}\Delta_{(1,2),(1\p,2\p)}
    + \tilde{\mcC}^{XW}_{L_1}\tilde{\mcC}^{YZ}_{L_1\p}\Delta_{(1,2\p),(1\p,2)} 
  \,, 
  \label{4delta}
} 
where the quantity $\tilde{\mcC}$ is the observed angular power spectrum, 
and $\Delta_{(1,1\p),(2,2\p)}$, $\Delta_{(1,2),(1\p,2\p)}$ and 
$\Delta_{(1,2\p),(1\p,2)}$ denote 
\al{ 
  \Delta_{(1,1\p),(2,2\p)} &\equiv \delta_{L_1,L_1\p}\delta_{L_2,L\p_2}
    \delta_{M_1,-M\p_1}\delta_{M_2,-M\p_2} 
  \,,\\ 
  \Delta_{(1,2),(1\p,2\p)} &\equiv \delta_{L_1,L_2}\delta_{L\p_1,L\p_2}
    \delta_{M_1,M_2}\delta_{M\p_1,M\p_2} 
  \,,\\ 
  \Delta_{(1,2\p),(1\p,2)} &\equiv \delta_{L_1,L\p_2}\delta_{L\p_1,L_2}
    \delta_{M_1,M\p_2}\delta_{M\p_1,M_2} 
  \,.
} 
The other terms included in the four-point correlation are represented 
by the second term in Eq.(\ref{GC}). 
If the quantities, $\tilde{X}_{L_1,M_1},$, $\tilde{Y}_{L\p_1,M\p_1}$, 
$\tilde{Z}_{L_2,M_2}$ and $\tilde{W}_{L\p_2,M\p_2}$, are random Gaussian 
fields, the second term vanishes. 
From Eq.(\ref{GC}), the covariance is decomposed into two parts: 
\beq 
  \ave{(\hat{x}^{(\alpha)}_{\ell,m})^*\hat{x}_{\ell,m}^{(\beta)}} 
    = G_{\ell,m}^{x,(\alpha,\beta)} + C_{\ell,m}^{x,(\alpha,\beta)} 
  \,,
\eeq 
where we define the Gaussian and connected parts, 
$G_{\ell,m}^{x,(\alpha,\beta)}$ and $C_{\ell,m}^{x,(\alpha,\beta)}$, as 
\al{ 
  G_{\ell,m}^{x,(\alpha,\beta)} &=  
      \sum_{L_1,L\p_1}\sum_{L_2,L\p_2}\sum_{M_1,M\p_1}\sum_{M_2,M\p_2} 
      \Wjm{\ell}{L_1}{L\p_1}{-m}{M_1}{M\p_1} 
      \Wjm{\ell}{L_2}{L\p_2}{-m}{M_2}{M\p_2} 
    \notag \\ 
    &\qquad \times 
      (F_{\ell,L_1,L\p_1}^{x,(\alpha)})^*F_{\ell,L_2,L\p_2}^{x,(\beta)} 
      \ave{\tilde{X}_{L_1,M_1}^*\tilde{Y}_{L\p_1,M\p_1}^*
      \tilde{Z}_{L_2,M_2}\tilde{W}_{L\p_2,M\p_2}}\rom{G} 
  \,, 
  \label{appB:G}
  \\ 
  C_{\ell,m}^{x,(\alpha,\beta)} &=  
      \sum_{L_1,L\p_1}\sum_{L_2,L\p_2}\sum_{M_1,M\p_1}\sum_{M_2,M\p_2} 
      \Wjm{\ell}{L_1}{L\p_1}{-m}{M_1}{M\p_1} 
      \Wjm{\ell}{L_2}{L\p_2}{-m}{M_2}{M\p_2} 
    \notag \\ 
    &\qquad \times 
      (F_{\ell,L_1,L_1\p}^{x,(\alpha)})^*F_{\ell,L_2,L\p_2}^{x,(\beta)} 
      \ave{\tilde{X}_{L_1,M_1}^*\tilde{Y}_{L\p_1,M\p_1}^*
      \tilde{Z}_{L_2,M_2}\tilde{W}_{L\p_2,M\p_2}}\rom{C}
  \,.
  \label{appB:C}
}

Let us first compute the Gaussian part, $G_{\ell,m}^{x,(\alpha,\beta)}$. 
Substituting Eq.(\ref{4delta}) into Eq.(\ref{appB:G}), the Gaussian 
part of the covariance is given by 
\al{ 
  G_{\ell,m}^{x,(\alpha,\beta)} 
    &= \sum_{L_1,L\p_1}\sum_{L_2,L\p_2}\sum_{M_1,M\p_1}\sum_{M_2,M\p_2} 
      \Wjm{\ell}{L_1}{L\p_1}{-m}{M_1}{M\p_1} 
      \Wjm{\ell}{L_2}{L\p_2}{-m}{M_2}{M\p_2} 
    \notag \\ 
    &\qquad \times 
      (F_{\ell,L_1,L\p_1}^{x,(\alpha)})^*F_{\ell,L_2,L\p_2}^{x,(\beta)} 
      \Big\{\tilde{\mcC}^{XY}_{L_1}\tilde{\mcC}^{ZW}_{L\p_1}\Delta_{(1,1\p),(2,2\p)}
    \notag \\ 
    &\qquad\qquad\qquad\qquad\qquad 
      + \tilde{\mcC}^{XZ}_{L_1}\tilde{\mcC}^{YW}_{L_1\p}\Delta_{(1,2),(1\p,2\p)}
      + \tilde{\mcC}^{XW}_{L_1}\tilde{\mcC}^{YZ}_{L_1\p}\Delta_{(1,2\p),(1\p,2)}
    \Big\} 
  \,.
  \label{appB:variance1}
}
Using Eq.(\ref{Wigner:sum1}), the term proportional to 
$\Delta_{(1,1\p),(2,2\p)}$ gives $\delta_{\ell,0}$, 
and we neglect this term to consider $\ell>0$. 
Then, Eq.(\ref{appB:variance1}) becomes 
\al{ 
  G_{\ell,m}^{x,(\alpha,\beta)}
    &= \sum_{L_1,L\p_1}\sum_{L_2,L\p_2}\sum_{M_1,M\p_1}\sum_{M_2,M\p_2} 
      \Wjm{\ell}{L_1}{L\p_1}{-m}{M_1}{M\p_1} 
      \Wjm{\ell}{L_2}{L\p_2}{-m}{M_2}{M\p_2} 
    \notag \\ 
    &\qquad \times (F_{\ell,L_1,L\p_1}^{x,(\alpha)})^*F_{\ell,L_2,L\p_2}^{x,(\beta)}
    \{\tilde{\mcC}^{XZ}_{L_1}\tilde{\mcC}^{YW}_{L_1\p}\Delta_{(1,2),(1\p,2\p)}
      + \tilde{\mcC}^{XW}_{L_1}\tilde{\mcC}^{YZ}_{L_1\p}\Delta_{(1,2\p),(1\p,2)}
    \} 
    \notag \\ 
    &= \sum_{L_1,L_1}\sum_{M_1,M\p_1} (F_{\ell,L_1,L\p_1}^{x,\alpha})^* 
      \notag \\ 
      &\qquad\times \bigg\{
        F_{\ell,L_1,L\p_1}^{x,(\beta)}\tilde{\mcC}^{XZ}_{L_1}\tilde{\mcC}^{YW}_{L\p_1}
        \Wjm{\ell}{L_1}{L\p_1}{-m}{M_1}{M\p_1} 
        \Wjm{\ell}{L_1}{L\p_1}{-m}{M_1}{M\p_1} \notag \\ 
      &\qquad\quad + F_{\ell,L\p_1,L_1}^{x,(\beta)}\tilde{\mcC}^{XW}_{L_1}\tilde{\mcC}^{YZ}_{L\p_1}
        \Wjm{\ell}{L_1}{L\p_1}{-m}{M_1}{M\p_1} 
        \Wjm{\ell}{L\p_1}{L_1}{-m}{M\p_1}{M_1} \bigg\} 
  \,.
} 
Using Eq.(\ref{sym0}), the above equation reduces to 
\al{
  G_{\ell,m}^{x,(\alpha,\beta)}
    &= \sum_{L_1,L_1} (F_{\ell,L_1,L\p_1}^{x,(\alpha)})^*
      (F_{\ell,L_1,L\p_1}^{x,(\beta)}\tilde{\mcC}^{XZ}_{L_1}\tilde{\mcC}^{YW}_{L\p_1}
      + (-1)^{\ell+L_1+L\p_1}F_{\ell,L\p_1,L_1}^{x,(\beta)}\tilde{\mcC}^{XW}_{L_1}\tilde{\mcC}^{YZ}_{L\p_1})
      \notag \\ 
      &\qquad\times \sum_{M_1,M\p_1} \Wjm{\ell}{L_1}{L\p_1}{-m}{M_1}{M\p_1} 
      \Wjm{\ell}{L_1}{L\p_1}{-m}{M_1}{M\p_1} \notag \\ 
    &= \frac{1}{2\ell+1} 
      \sum_{L_1,L_1} (F_{\ell,L_1,L\p_1}^{x,(\alpha)})^* \notag \\ 
      &\qquad\times \left(F_{\ell,L_1,L\p_1}^{x,(\beta)} \tilde{\mcC}^{XZ}_{L_1}\tilde{\mcC}^{YW}_{L\p_1}
      + F_{\ell,L\p_1,L_1}^{x,(\beta)} (-1)^{\ell+L_1+L\p_1}
      \tilde{\mcC}^{XW}_{L_1}\tilde{\mcC}^{YZ}_{L\p_1}\right) 
  \,.
  \label{appB:g}
}
Note that we use (\ref{Wigner:sum2}) from the first to the last equation. 

Next we compute the connected part of the covariance (\ref{appB:C}). 
Even if the primordial CMB temperature and polarization anisotropies 
are random Gaussian fields, the connected part of the four-point correlations 
is arising from the secondary effects such as the weak lensing. 
Assuming that the connected part is arising from the lensing effect, 
the connected part of the four-point correlations is given as 
\al{ 
  &\ave{\tilde{X}^*_{L_1,M_1}\tilde{Y}^*_{L\p_1,M\p_1}
    \tilde{Z}_{L_2,M_2} \tilde{W}_{L\p_2,M\p_2}}\rom{C} \notag \\ 
  &\qquad\simeq 
    \sum_{\ell\pp,m\pp}
    \Wjm{\ell\pp}{L_1}{L\p_1}{-m\pp}{M_1}{M\p_1}
    \Wjm{\ell\pp}{L_2}{L\p_2}{-m\pp}{M_2}{M\p_2} 
    \sum_{x\p}(f^{x\p,(\alpha)}_{\ell\pp,L_1,L\p_1})^*
    f^{x\p,(\beta)}_{\ell\pp,L_2,L\p_2}C_{\ell\pp}^{x\p x\p}
  \,. 
  \label{connect}
} 
Other terms included in the connected part of four-point correlation, 
such as the non-Gaussian terms introduced in Ref.\cite{Cooray:2002py}, 
induce additional noise term in Eq.(\ref{3:covar}), 
but the terms may be an order of magnitude smaller than 
$G_{\ell}^{x,(\alpha,\beta)}$ \cite{Kesden:2003cc,Cooray:2002py,Hanson:2010rp}. 
Substituting Eq.(\ref{connect}) into Eq.(\ref{appB:C}), the connected 
part of the covariance is rewritten as 
\al{ 
  C_{\ell,m}^{x,(\alpha,\beta)} &=  
    \sum_{L_1,L\p_1}\sum_{L_2,L\p_2}\sum_{M_1,M\p_1}\sum_{M_2,M\p_2} 
    \Wjm{\ell}{L_1}{L\p_1}{-m}{M_1}{M\p_1} 
    \Wjm{\ell}{L_2}{L\p_2}{-m}{M_2}{M\p_2} \notag \\ 
  &\qquad \times 
    (F_{\ell,L_1,L\p_1}^{x,(\alpha)})^*F_{\ell,L_2,L\p_2}^{x,(\beta)} 
    \sum_{\ell\pp,m\pp} 
    \Wjm{\ell\pp}{L_1}{L\p_1}{-m\pp}{M_1}{M\p_1}
    \Wjm{\ell\pp}{L_2}{L\p_2}{-m\pp}{M_2}{M\p_2} 
  \notag \\ 
  &\qquad\qquad\qquad\qquad\qquad\times 
    \sum_{x\p}(f^{x\p,(\alpha)}_{\ell\pp,L_1,L\p_1})^*
    f^{x\p,(\beta)}_{\ell\pp,L_2,L\p_2}C_{\ell\pp}^{x\p x\p} 
  \,. 
}
With Eq.(\ref{Wigner:sum2}), the above equation is rewritten as 
\al{
  C_{\ell,m}^{x,(\alpha,\beta)} &=  
    \sum_{L_1,L\p_1}\sum_{L_2,L\p_2}\sum_{\ell\pp,m\pp}\sum_{x\p} 
    \frac{\delta_{\ell,\ell\pp}\delta_{m,m\pp}}{2\ell+1}
    \frac{\delta_{\ell,\ell\pp}\delta_{m,m\pp}}{2\ell+1}
    \notag \\ 
  &\qquad \times 
    (F_{\ell,L_1,L_1\p}^{x,(\alpha)})^*F_{\ell,L_2,L_2\p}^{x,(\beta)} 
    (f^{x\p,(\alpha)}_{\ell\pp,L_1,L\p_1})^*
    f^{x\p,(\beta)}_{\ell\pp,L_2,L\p_2}C_{\ell}^{x\p,x\p} 
  \notag \\
  &= \sum_{x\p}\Big([F^x,f^{x\p}]_{\ell}^{(\alpha)}\Big)^*
    [F^x,f^{x\p}]_{\ell}^{(\beta)}C_{\ell\p}^{x\p x\p} 
  \,. 
} 
Using Eq.(\ref{3:unbias}), we obtain 
\al{
  C_{\ell,m}^{x,(\alpha,\beta)} &= C_{\ell\p}^{xx} 
  \,.
  \label{appB:c}
}
Finally, substituting the resultant form of the Gaussian (\ref{appB:g}) 
and connected parts (\ref{appB:c}) into Eq.(\ref{appB:n-cov}), we obtain 
Eq.(\ref{3:covar}).

\section{
Flat-sky limit
} 
\label{appC}

In this appendix, using Eqs.(\ref{abl-alm})-(\ref{exp-Ylm}), we derive 
Eqs.(\ref{flat:g-func}), (\ref{flat:N}) and the noise cross-spectrum 
in the flat-sky limit (\ref{flat:covar}). 

We first consider the expression for 
$\mcT_{\bl,\bL,\bL\p}g^{x,(\alpha)}_{\ell,L,L\p}$ (\ref{flat:g-func}) 
in the flat-sky limit. 
From Eq.(\ref{full:g-func}), we obtain 
\al{ 
  \mcT_{\bl,\bL,\bL\p}g^{x,(\alpha)}_{\ell,L,L\p}
    &= \frac{\tilde{\mcC}^{XX}_{L\p}\tilde{\mcC}^{YY}_L
      \mcT_{\bl,\bL,\bL\p}(f^{x,(\alpha)}_{\ell,L,L\p})^* 
      - \tilde{\mcC}_{L}^{XY}\tilde{\mcC}_{L\p}^{XY}
      \mcT_{\bl,\bL\p,\bL}(f^{x,(\alpha)}_{\ell,L\p,L})^*}
    {\tilde{\mcC}^{XX}_{L\p}\tilde{\mcC}^{YY}_L
     \tilde{\mcC}^{XX}_{L}\tilde{\mcC}^{YY}_{L\p}
      - (\tilde{\mcC}^{XY}_L\tilde{\mcC}^{XY}_{L\p})^2} 
  \,. 
  \label{Tg-func}
} 
Note that, we use 
\beq 
  \mcT_{\bl,\bL,\bL\p}=(-1)^{\ell+L+L\p}\mcT_{\bl,\bL\p,\bL} 
  \,.
\eeq 
Then, we need to compute the quantity, 
$(\mcT_{\bl,\bL,\bL\p})^*f^{x,(\alpha)}_{\ell,L,L\p}$. 
As shown in Table.\ref{table:f}, this quantity includes 
$(\mcT_{\bl,\bL,\bL\p})^*\,_s\mcS^{x}_{L\p,\ell,L}$ and 
$(\mcT_{\bl,\bL,\bL\p})^*\,_s\mcS^{x}_{L,\ell,L\p}$ where 
$s=0$ or $\pm2$. 
From Eqs.(\ref{mcSg}), (\ref{mcSc}), (\ref{2mcSg}), and (\ref{2mcSc}), 
we obtain 
\al{
  (\mcT_{\bl,\bL,\bL\p})^*\,_s\mcS^x_{L\p,\ell,L} 
    &= \left(\frac{(2L+1)(2L\p+1)}{4\pi(2\ell+1)(LL\p)^2}\right)^{1/2}
      \sum_{m,M,M\p}e^{-im\varphi_{\ell}}e^{iM\varphi_L}
      e^{iM\p\varphi_{L\p}}(-1)^{m+M\p}
    \notag \\ 
    &\qquad\times
    i^{m-M-M\p}\int d^2\hatn \,_s\mcY^*_{L\p,-M\p}(\hatn)
    [\bm{\nabla}\,_0\mcY_{\ell,-m}(\hatn)]\odot_x
    [\bm{\nabla}\,_s\mcY_{L,M}(\hatn)] 
  \,,
  \label{appC:TS}
} 
where, for arbitrary two vectors, $\bm{a}$ and $\bm{b}$, we define the 
products, $\odot_{\grad}$ and $\odot_{\curl}$, as 
\beq 
  \bm{a}\odot_{\grad}\bm{b} \equiv \bm{a}\cdot\bm{b}, \qquad 
  \bm{a}\odot_{\curl}\bm{b} \equiv (\star\bm{a})\cdot\bm{b} 
    = -(\star\bm{b})\cdot\bm{a}
  \,.
\eeq 
Note here that, the right-hand side of Eqs.(\ref{mcSg}) and 
(\ref{2mcSg}) is a real number, since the left-had side of these 
equations is a real number. 
Similarly, the right-hand sides of Eqs.(\ref{mcSc}) and 
(\ref{2mcSc}) is a purely imaginary number. 
Thus, Eq.(\ref{appC:TS}) is rewritten as 
\al{
  (\mcT_{\bl,\bL,\bL\p})^*\,_s\mcS^x_{L\p,\ell,L} 
  &= \left(\frac{(2L+1)(2L\p+1)}{4\pi(2\ell+1)(LL\p)^2}\right)^{1/2}
      \sum_{m,M,M\p}e^{-im\varphi_{\ell}}e^{iM\varphi_L}
      e^{iM\p\varphi_{L\p}}(-1)^{m+M\p}
    \notag \\ 
    &\qquad\times
    i^{m-M-M\p}\int d^2\hatn \,_s\mcY_{L\p,-M\p}(\hatn)
    [\bm{\nabla}\,_s\mcY^*_{L,M}(\hatn)] 
    \odot_x[\bm{\nabla}\,_0\mcY^*_{\ell,-m}(\hatn)]
  \notag \\
  &= \left(\frac{\ell(2L+1)(2L\p+1)}{2(2\ell+1)LL\p}\right)^{1/2}
      \notag \\ 
    &\qquad \times 
      \sum_{m,M,M\p}\sqrt{\frac{2\pi}{\ell}}i^me^{-im\varphi_{\ell}}
      \sqrt{\frac{2\pi}{L}}i^{-M}e^{iM\varphi_L}
      \sqrt{\frac{2\pi}{L\p}}i^{-M\p}e^{iM\p\varphi_{L\p}} 
      \notag \\ 
    &\qquad\times
    \Int{\hatn} \,_s\mcY^*_{L\p,M\p}(\hatn)
    [\bm{\nabla}\,_s\mcY^*_{L,M}(\hatn)] 
    \odot_x[\bm{\nabla}\,_0\mcY_{\ell,m}(\hatn)]
  \,. 
} 
Using Eq.(\ref{exp-Ylm}) and assuming $\ell,L,L\p\gg1$, the above 
equation reduces to 
\al{
  (\mcT_{\bl,\bL,\bL\p})^*\,_s\mcS^x_{L\p,\ell,L} 
    &\simeq e^{si(\varphi_{L\p}-\varphi_L)}
    \bL\odot_x\bl \int \frac{d^2\hatn}{(2\pi)^2} 
    e^{i(\bl-\bL-\bL\p)\cdot\hatn} 
  \,.
}
From Eq.(\ref{delta}), we obtain 
\al{
  (\mcT_{\bl,\bL,\bL\p})^*\,_s\mcS^{x}_{L\p,\ell,L} 
    &\simeq \delta_{\bL+\bL\p-\bl}\,\bL\odot_x\bl \times 
    \begin{cases}
      1 & (s=0) \\
      \cos 2\varphi_{L,L\p} & (s=\oplus) \\ 
      -\sin 2\varphi_{L,L\p} & (s=\ominus) 
    \end{cases} \label{flat:s=0+-} 
  \,,
}
with $\varphi_{L,L\p}=\varphi_L-\varphi_{L\p}$. 
Similarly, the flat-sky counterpart of 
$(\mcT_{\bl,\bL,\bL\p})^*\,_s\mcS^{x}_{L,\ell,L\p}$ is obtained by 
interchanging $\bL$ and $\bL\p$ in Eq.(\ref{flat:s=0+-}) if $\ell+L+L\p$ 
is an even integer. If $\ell+L+L\p$ is an odd integer, we further 
multiply it by minus sign. From Eq.(\ref{flat:s=0+-}), we can define 
the following quantity: 
\al{
  \ol{f}^{x,(\alpha)}_{\bl,\bL,\bL\p}\delta_{\bL+\bL\p-\bl}
    &=(\mcT_{\bl,\bL,\bL\p})^*f^{x,(\alpha)}_{\ell,L,L\p} 
  \,.
  \label{flat:f} 
}
The functional form of $\ol{f}^{x,(\alpha)}_{\bl,\bL,\bL\p}$ is 
summarized in Table.\ref{table:olf} for each $x$ and $\alpha$. 
Substituting Eq.(\ref{flat:f}) into Eq.(\ref{Tg-func}), we obtain the 
following expression: 
\al{ 
  \mcT_{\bl,\bL,\bL\p}g^{x,(\alpha)}_{\ell,L,L\p}
    &= \delta_{\bL+\bL\p-\bl}\frac{\tilde{\mcC}^{XX}_{L\p}\tilde{\mcC}^{YY}_L
      (\ol{f}^{x,(\alpha)}_{\bl,\bL,\bL\p})^* 
      - \tilde{\mcC}_{L}^{XY}\tilde{\mcC}_{L\p}^{XY}
      (\ol{f}^{x,(\alpha)}_{\bl,\bL\p,\bL})^*}
    {\tilde{\mcC}^{XX}_{L\p}\tilde{\mcC}^{YY}_L
     \tilde{\mcC}^{XX}_{L}\tilde{\mcC}^{YY}_{L\p}
      - (\tilde{\mcC}^{XY}_L\tilde{\mcC}^{XY}_{L\p})^2} 
    \notag \\
    &= \delta_{\bL+\bL\p-\bl}\ol{g}^{x,(\alpha)}_{\bl,\bL,\bL\p}
  \,, 
  \label{Tg-func2}
} 
with the quantity, $\ol{g}^{x,(\alpha)}_{\bl,\bL,\bL\p}$, being 
\al{
  \ol{g}^{x,(\alpha)}_{\bl,\bL,\bL\p}
    &= \frac{\tilde{\mcC}^{XX}_{L\p}\tilde{\mcC}^{YY}_L
      (\ol{f}^{x,(\alpha)}_{\bl,\bL,\bL\p})^* 
      - \tilde{\mcC}_{L}^{XY}\tilde{\mcC}_{L\p}^{XY}
      (\ol{f}^{x,(\alpha)}_{\bl,\bL\p,\bL})^*}
    {\tilde{\mcC}^{XX}_{L\p}\tilde{\mcC}^{YY}_L
     \tilde{\mcC}^{XX}_{L}\tilde{\mcC}^{YY}_{L\p}
      - (\tilde{\mcC}^{XY}_L\tilde{\mcC}^{XY}_{L\p})^2} 
  \,.
  \label{appC:g-func}
}

Next, we consider the flat-sky counterpart of Eq.(\ref{3:covar}), and 
show Eqs.(\ref{flat:N}) and (\ref{flat:covar}). 
Using Eq.(\ref{Wigner:sum2}), the covariance (\ref{3:covar}) is 
rewritten as 
\al{
  N_{\ell}^{x,(\alpha,\beta)} 
    &= \sum_{m,M,M\p}\sum_{m\p,M\pp,M\ppp}
       \frac{(-1)^{m+m\p}\delta_{m,m\p}\delta_{M,M\pp}\delta_{M\p,M\ppp}}{2\ell+1}
       \Wjm{\ell}{L}{L\p}{-m}{M}{M\p}\Wjm{\ell}{L}{L\p}{-m\p}{M\pp}{M\ppp} 
       \notag \\
    &\qquad\times\sum_{L,L\p} (F_{\ell,L,L\p}^{x,(\alpha)})^*
      \left(F_{\ell,L,L\p}^{x,(\beta)}\tilde{\mcC}_L^{XX\p}\tilde{\mcC}_{L\p}^{YY\p}
      + F_{\ell,L\p,L}^{x,(\beta)} (-1)^{\ell+L+L\p}
      \tilde{\mcC}_L^{XY\p}\tilde{\mcC}_{L\p}^{X\p Y}\right) 
      \notag \\ 
    &= \sum_{m,M,M\p}\sum_{m\p,M\pp,M\ppp}\frac{(-1)^{m+m\p}}{2\ell+1} 
      \Wjm{\ell}{L}{L\p}{-m}{M}{M\p}\Wjm{\ell}{L}{L\p}{-m\p}{M\pp}{M\ppp} 
      \notag \\
    &\qquad\times 
      \int\frac{d\varphi_{\ell}}{2\pi}e^{-i(m-m\p)\varphi}
      \int\frac{d\varphi_L}{2\pi}e^{-i(M-M\pp)\varphi_L}
      \int\frac{d\varphi_{L\p}}{2\pi}e^{-i(M\p-M\ppp)\varphi_{L\p}}
      \notag \\
    &\qquad\times\sum_{L,L\p} (F_{\ell,L,L\p}^{x,(\alpha)})^*
      \left(F_{\ell,L,L\p}^{x,(\beta)}\tilde{\mcC}_L^{XX\p}\tilde{\mcC}_{L\p}^{YY\p}
      + F_{\ell,L\p,L}^{x,(\beta)} (-1)^{\ell+L+L\p}
      \tilde{\mcC}_L^{XY\p}\tilde{\mcC}_{L\p}^{X\p Y}\right) 
  \,,
  \label{appC:N1}
}
where, for arbitrary integers, $M_1$ and $M_2$, we use the following 
equation: 
\beq 
  \delta_{M_1,M_2} = \int\frac{d\varphi}{2\pi}e^{-i(M_1-M_2)\varphi} 
  \,. 
\eeq 
With the quantity defined in Eq.(\ref{T-func}), the expression for 
the covariance (\ref{appC:N1}) is rewritten as 
\al{
  N_{\ell}^{x,(\alpha,\beta)} 
    &= \int\frac{d\varphi_{\ell}}{2\pi}\sum_{L,L\p}
      \left(\frac{(2L+1)(2L\p+1)}{4\pi(LL\p)^2}\right)^{-1}
      \int\frac{d\varphi_{L}}{2\pi}\int\frac{d\varphi_{L\p}}{2\pi}
      (\mcT_{\bl,\bL,\bL\p})^*\mcT_{\bl,\bL,\bL\p} 
      \notag \\
    &\qquad\times (F_{\ell,L,L\p}^{x,(\alpha)})^*
      \left(F_{\ell,L,L\p}^{x,(\beta)}\tilde{\mcC}_L^{XX\p}\tilde{\mcC}_{L\p}^{YY\p}
      + F_{\ell,L\p,L}^{x,(\beta)} (-1)^{\ell+L+L\p}
      \tilde{\mcC}_L^{XY\p}\tilde{\mcC}_{L\p}^{X\p Y}\right) 
  \,.
  \label{appC:N2}
}

In the flat-sky limit, we can define the following quantity: 
\al{
  \delta_{\bL+\bL\p-\bl} \ol{F}_{\bl,\bL\p,\bL}^{x,(\beta)} 
    &\simeq \mcT_{\bl,\bL,\bL\p}F_{\ell,L,L\p}^{x,(\beta)}
  \,. 
  \label{F-delta}
}
This is because, from Eqs.(\ref{Tg-func2}), the right-hand side of the 
above equation is proportional to the delta function, 
$\delta_{\bL+\bL\p-\bl}$. 
Using Eq.(\ref{F-delta}), and $\delta_{\bm{0}}=1/\pi$, and assuming 
$\ell,L,L\p\gg 1$, Eq.(\ref{appC:N2}) becomes 
\al{
  N_{\ell}^{x,(\alpha,\beta)} 
    &\simeq \int\frac{d\varphi_{\ell}}{2\pi}\sum_{L,L\p}LL\p 
      \int\frac{d\varphi_{L}}{2\pi}\int\frac{d\varphi_{L\p}}{2\pi} \delta_{\bL+\bL\p-\bl}
      \notag \\
    &\qquad\times (\ol{F}_{\bl,\bL,\bL\p}^{x,(\alpha)})^*
      \left(\ol{F}_{\bl,\bL,\bL\p}^{x,(\beta)}\tilde{\mcC}_L^{XX\p}\tilde{\mcC}_{L\p}^{YY\p}
      + \ol{F}_{\bl,\bL\p,\bL}^{x,(\beta)}
      \tilde{\mcC}_L^{XY\p}\tilde{\mcC}_{L\p}^{X\p Y}\right) 
  \,.
}
In the right-hand side, we can choose two-dimensional 
coordinate system for the variables of integration, $\bL$ and $\bL\p$, 
so that the integrand of $\varphi_{\ell}$ does not 
depend on $\varphi_{\ell}$. 
Then, the above equation reduces to 
\al{
  \ol{N}_{\ell}^{x,(\alpha,\beta)} &= \Int{\bL}\int d^2\bL\p \delta_{\bL+\bL\p-\bl} 
    (\ol{F}^{x,(\alpha)}_{\bl,\bL,\bL\p})^* \notag \\ 
    &\qquad\times 
    \left(\ol{F}^{x,(\beta)}_{\bl,\bL,\bL\p}\mcC_L^{XX\p}\mcC_{L\p}^{YY\p}
    +\ol{F}^{x,(\beta)}_{\bl,\bL\p,\bL}\mcC_L^{XY\p}\mcC_{L\p}^{X\p Y}\right)
  \,. 
  \label{appC:cov}
} 
The noise in the flat-sky limit, $\ol{N}_{\ell}^{x,(\alpha)}$, is obtained 
by $\alpha=\beta$ in the above equation: 
\al{
  \ol{N}_{\ell}^{x,(\alpha)} &= \Int{\bL}\int d^2\bL\p \delta_{\bL+\bL\p-\bl} 
    (\ol{F}^{x,(\alpha)}_{\bl,\bL,\bL\p})^* \notag \\ 
    &\qquad\times 
    \left(\ol{F}^{x,(\alpha)}_{\bl,\bL,\bL\p}\mcC_L^{XX\p}\mcC_{L\p}^{YY\p}
    +\ol{F}^{x,(\alpha)}_{\bl,\bL\p,\bL}\mcC_L^{XY\p}\mcC_{L\p}^{X\p Y}\right)
  \,.
  \label{appC:N3}
} 
Note that the quantity $\ol{F}^{x,(\alpha)}_{\bl,\bL,\bL\p}$ is described 
by 
\beq 
  \ol{F}^{x,(\alpha)}_{\bl,\bL,\bL\p} 
    = \ol{N}_{\ell}^{x,(\alpha)}\ol{g}^{x,(\alpha)}_{\bl,\bL,\bL\p}
  \,.
  \label{appC:F}
\eeq 
Substituting Eq.(\ref{appC:F}) into Eq.(\ref{appC:N3}), we obtain the 
expression for the noise spectrum in the flat-sky limit:  
\al{
  \ol{N}_{\ell}^{x,(\alpha)} 
    &= \left\{\Int{\bL}\int d^2\bL\p \delta_{\bL+\bL\p-\bl} 
    \ol{f}^{x,(\alpha)}_{\bl,\bL,\bL\p}\ol{g}^{x,(\alpha)}_{\bl,\bL,\bL\p}\right\}^{-1}
  \,.
  \label{appC:N}
} 

\bibliographystyle{mybst}
\bibliography{cite}

\end{document}